 \documentclass[preprint2,longabstract]{aastex}

\usepackage{algorithmic}
\usepackage{algorithm}

\def\la{\lower.5ex\hbox{$\; \buildrel < \over \sim \;$}}
\def\ga{\lower.5ex\hbox{$\; \buildrel > \over \sim \;$}}

\newcommand{\HI}{\mbox{H{\sc i}}}

\newcommand{\masq}{\mbox{mag~arcsec$^{-2}$}}
\newcommand{\am}[2]{$#1'\,\hspace{-1.7mm}.\hspace{.1mm}#2$} 
\newcommand{\as}[2]{$#1''\,\hspace{-1.7mm}.\hspace{.1mm}#2$} 

\shorttitle{Sustaining star formation rates in spirals}
\shortauthors{Vollmer \& Leroy}

\begin{document}


\title{Simultaneous multi-band detection of Low Surface Brightness galaxies with Markovian modelling\thanks{The presented software can be downloaded at http://lsiit-miv.u-strasbg.fr/paseo/LSBdetection.php.}}


\author{B.~Vollmer}
\affil{CDS, Observatoire astronomique, UMR 7550, 11 rue de l'universit\'e, 67000 Strasbourg, France}\email{Bernd.Vollmer@astro.unistra.fr}

\author{B.~Perret\footnote{present address: Universit\'e Paris-Est, Laboratoire d'Informatique Gaspard-Monge, Equipe A3SI, ESIEE Paris}, 
M.~Petremand, F.~Lavigne, Ch.~Collet}
\affil{LSIIT, Universit\'e de Strasbourg, 7, rue Ren\'e Descartes, 67084 Strasbourg, France}

\author{W.~van Driel}
\affil{GEPI, Observatoire de Paris, CNRS, Universit\'e Paris Diderot, 5 place Jules Janssen, 92190 Meudon, France}

\author{F.~Bonnarel, M.~Louys}
\affil{CDS, Observatoire astronomique, UMR 7550, 11 rue de l'universit\'e, 67000 Strasbourg, France}

\author{S.~Sabatini}
\affil{INAF/IASF-Roma, via Fosso de Cavaliere 100, 00133 Roma, Italy}

\and

\author{L.~A.~ MacArthur}
\affil{Herzberg Institute of Astrophysics, National Research Council of Canada,
Victoria, BC V9E 2E7, Canada}

%% are identified by the \altaffilmark after each name. Specify alternate


\begin{abstract}
We present to the astronomical community an algorithm for the detection of Low Surface Brightness (LSB) galaxies in images, called MARSIAA (MARkovian Software for Image Analysis in Astronomy), which is based on multi-scale Markovian modeling. MARSIAA can be applied simultaneously to different bands. It segments an image into a user-defined number of classes, according to their surface brightness and surroundings -- typically, one or two classes contain the LSB structures. We have developed an algorithm, called DetectLSB, which allows the efficient identification of LSB galaxies from among the candidate sources selected by MARSIAA. The application of the method to two and three bands simultaneously was tested on simulated images. Based on our tests we are confident that we can detect LSB galaxies down to a central surface brightness level of only 1.5 times the standard deviation from  the mean pixel value in the image background. To assess the  robustness of our method, the method was applied to a set of 18 B and I band images (covering 1.3 square degrees in total) of the Virgo cluster to which Sabatini et al. (2003, 2005) previously applied a matched-filters dwarf LSB galaxy search algorithm. We have detected all 20 objects from the Sabatini et al. catalog which we could classify by eye as bona fide LSB galaxies. Our method has also detected 4 additional Virgo cluster LSB galaxy candidates undetected by Sabatini et al. 
To further assess the completeness of the results of our method, both MARSIAA, SExtractor, and DetectLSB were applied to search for (i) mock Virgo LSB galaxies inserted into a set of deep Next Generation Virgo Survey (NGVS) gri-band subimages and (ii) Virgo LSB galaxies identified by eye in a full set of NGVS square degree gri images. MARSIAA/DetectLSB recovered $\sim 20$\,\% more mock LSB galaxies and $\sim 40$\,\% more LSB galaxies  identified by eye than SExtractor/DetectLSB.
With a $90$\,\% fraction of false positives from an entirely unsupervised pipeline, a completeness of $90$\,\% is reached for sources with $r_{\rm e} > 3''$ at a mean surface brightness level of $\mu_{\rm g}=27.7$~\masq and a central surface brightness of $\mu^{0}_{\rm g}=26.7$~\masq.
About $10$\,\% of the false positives are artifacts, the rest being background galaxies.
We have found our proposed Markovian LSB galaxy detection method to be complementary to the application of matched filters and an optimized use of SExtractor, and to have the following advantages: it is scale-free, can be applied simultaneously to several bands, and is well adapted for crowded regions  on the sky.
\end{abstract}


\keywords{Methods: data analysis -- Techniques: image processing -- Galaxies: fundamental parameters -- Galaxies: clusters individual: Virgo}

%% and \citet commands to identify citations. The citations are

%% their objects with \objectname{} or \object{}. Each macro takes the
%% differs from what is to be printed in the paper. The text appearing 
%% Otherwise, LaTeX will issue an error. 

\section{Introduction \label{sec:introduction}}

The determination of the baryonic matter content and distribution is a crucial issue in the present era of precision cosmology, in which the detectability of galaxies in optical data plays a significant role. 
Compared to the selection effects affecting ``normal'' High Surface Brightness (HSB) galaxies, whose examples fill most of the pages of classics like The Hubble Atlas of Galaxies (Sandage 1961), those concerning the detection and identification of Low Surface Brightness (LSB) galaxies are much more challenging, due to their very nature (e.g., Disney 1976) -- a rule-of-thumb working definition of LSB galaxies is objects with a central surface brightness at least one magnitude fainter than the dark night sky. 
Important  issues in which  LSB galaxies play a key role are the slope of the faint end of the luminosity function and the extension of the color-magnitude relation to the faint luminosity regime (see, e.g., Ferrarese et al. 2012). Both subjects give strong constraints on the global, macroscopic processes that governed the assembly of baryons within merging dark matter halos. In addition, constructing and interpreting scaling relations for LSB galaxies will increase our knowledge on galaxy formation and evolution.
% Allen & Shu 1979; Disney & Phillips 1983 ; Davies 1990

As a class of objects, a significant part of the LSB galaxies population had been missed for more than two decades, before they became the object of a series of studies in the 80's (e.g. Bothun 1985; 1986; Davies et al. 1988), triggered by, e.g. the ability of CCD cameras to detect galaxies of considerably lower surface brightness than photographic plates, and the discovery that most of the LSBs then identified were quite rich in \HI\ gas and therefore easy to detect and map with radio telescopes.
After the first studies, a consensus seemed to emerge in the literature of the mid-90's that LSB galaxies are dwarfish, blue, and \HI-rich, indicating that their potential contribution to the baryonic matter reservoir may be relatively minor (e.g., Impey \& Bothun 1997; Bothun et al. 1997, and references therein). However, further studies started to reveal significant numbers of LSBs that did not fall within these narrow confines (e.g., O'Neil et al. 2000, 2004, Sabatini et al. 2003, Roberts et al. 2004). 
The observed diversity of LSBs made van den Bergh (1998) suggest their classification into at least three classes: "monsters" with sizes comparable to those of galaxy cluster cores (like Malin 1), galaxies with sizes like those of most "normal" (HSB) spirals, and dwarf irregular/spiral/elliptical galaxies. Only the latter class incorporates the traditionally recognized LSB galaxies.
Even though LSBs are now well established as a genuine class of galaxies with often 'extreme' properties, considerable uncertainty remains as to the scope of their physical properties and their number density in the local Universe. We have clearly not yet fully mapped out the parameter space they occupy.
Furthermore, we should not forget that our concept of Low Surface Brightness galaxies remains limited by our ability to detect extended objects just above the current noise limit of images -- the increased sensitivity of CCDs over photographic plates and the development of new detection algorithms have unlocked a wealth of hitherto undetectable objects, but it should be emphasized that there may well still be large numbers of extremely LSB galaxies out there beyond our present detection limits. Contemporary detection limits have led to contemporary working definition of LSB galaxies, such as the commonly used classification based on an extrapolated central blue disk surface brightness 
$\mu_{B}^0$$\geq$ 23 \masq\ (before or after deprojection to face-on), which basically means a disk galaxy with a $\mu_{B}^0$ about at least one \masq\ fainter than the typical mean value for what are now referred to as HSB spiral galaxies, i.e., the remarkably tight range of 21.7$\pm$0.3 \masq\ found by Freeman (1970).

Some of the specific issues regarding the problem of LSB galaxy detection are:
\begin{itemize}
\item
a large fraction, if not the entire, LSB galaxy is buried in the noise, where the noise is defined as the {\it standard deviation} of the pixel values from the image background; 
\footnote{This is different from the definition commonly used in image processing, i.e., the {\it variance} of the pixel values from the image background.}
\item
the morphology of LSB galaxies covers a very broad range, from round (elliptical galaxies) and elliptical (spiral galaxies) to irregular (irregular galaxies). Most spiral galaxies have a bulge and a disk component, and it is well possible that only the high surface brightness bulge (round) is detected while the much dimmer disk remains below the threshold level -- these objects are thus misclassified as dwarf ellipticals and considered as low-mass, gas-free objects. The seminal case of the giant LSB spiral Malin 1 is an example of this (Bothun et al. 1987);
\item
substructure is often present, in the form of spiral arms and distinct irregular star formation regions;
\item
since LSB galaxies span a wide range of colors, their surface brightness distribution, i.e., morphology, can be quite different in different bands. Whereas ellipticals are brightest in the red bands, irregulars are brightest in the blue bands;
\item
the area covered by the LSB galaxy can be polluted by multiple foreground stars or background galaxies of different magnitudes.
\end{itemize}

In practice, different algorithms have been used to detect LSB galaxies just above the noise level in optical and near-infrared images, such as:

SExtractor: The Source EXtractor software (Bertin \& Arnouts 1996) which is widely used to detect, deblend, measure and classify sources from images can also be used in a non-standard, adjusted way to search for LSB galaxies, e.g. with an extremely low cut at $1.2 \sigma$ and at least 200 connected pixels (Sect.~\ref{sec:sextractor}). 

Ring filtering: the ring filter is defined as a median filter which assigns weight only to selected pixels in an annulus. The filter has a sharply-defined scale-length; that is, all objects with a scale-size less than the radius of the ring are filtered and replaced by the local background level. It provides a fast, simple, and intuitive method to remove small-scale objects (independent of morphology) from an image, leaving behind the large-scale objects and overall light gradients (Secker 1995).

Matched filters: Sabatini et al. (2003) developed a method based on image convolution with a set of matched filters with different sizes that are optimized to enhance faint structures with scales that match the filter scale. The results per filter are combined into a single significance image in which objects of all different sizes are emphasized. 
With this method a complete sample of galaxies was selected down to a central surface brightness level of $\mu_{\rm B}$=25.3~\masq. The selection of LSB Virgo cluster member galaxies was based on morphological criteria, i.e., scale-lengths between 2 and 10~arcsec. 

Alternative detection methods, such as wavelet, curvelet or shapelet (see Crouse et al. 1998; Romberg et al. 2001; Murtagh et al. 2005; Starck \& Murtagh 2006; Chan et al. 2008), may be used to detect and de-noise low surface brightness structures.
Although some of these methods have been applied to observations obtained in two or more bands, we are not aware of any LSB galaxy detection algorithm which has been applied simultaneously to multiple bands.

Matched filters have also been applied to multiple bands in the mm wavelength range to extract cluster catalogs from Sunyaev-Zel'dovich (SZ) effect surveys (Melin et al. 2006). 
Pires et al. (2006) introduced an unsupervised analysis on simulated multi-band data to separate the SZ signal
(blind separation of the different components using an Independent Component Analysis (ICA) method), followed by non-linear wavelet 
filtering and application of SExtractor. Starck et al. (2006) used the multiscale entropy concept based on wavelets and the False Discovery Rate (FDR) to robustly reconstruct weak lensing maps. The FDR procedure provides the means to adaptively control the fraction of false discoveries over total discoveries. 

The latter methods rely on pre-defined morphologies for the source detection, whereas our aim is to provide a robust, scale-free, and morphology-independent LSB galaxy detection algorithm for structures which are deeply embedded in the pixel noise. In addition, the possibility to extend a method to multi-band data is required to optimize the detection process.

This paper represents a pilot study in which we apply our algorithm to (i) INT B and I band images of the Virgo cluster, for which Sabatini et al. (2003, 2005) have already obtained LSB galaxy catalogs using matched filters and (ii) to a set of NGVS (Next Generation Virgo Survey, Ferrarese et al. 2012) gri band subimages in which mock Virgo LSB galaxies were inserted, and (iii) to a full set of NGVS square degree gri images in which LSB galaxies were identified by eye. The challenge to our scale-free Markovian detection method is to show 
\begin{itemize}
\item
the robustness of the detection of LSB galaxies of different morphologies and with different degrees of foreground point source pollutions and source confusion;
\item
that its simultaneous application to two or more broad-band filters significantly improves the detection rate compared to its consecutive application to each filter. 
\end{itemize}
Our Markovian segmentation algorithm is embedded into an entirely unsupervised pipeline for the detection and identification of LSB galaxy candidates in 
multiband optical images. We aim at providing the most complete list of LSB galaxy candidates with a minimum number of false positives without human
intervention. Our method is not intended to provide definite and precise sizes or magnitudes which is beyond the scope of this work.  
The single-band application of the algorithm is described in Laferte et al. (2000), and the application to multiple bands in Provost et al. (2004). The latter study  determined water depth maps from SPOT satellite images. The novelty of our work lies in the use of the algorithm for LSB galaxy detection in astronomical images. For an alternative bayesian inference method for multiband image segmentation see Murtagh et al. (2005).

This paper is structured as follows: the principles of the multi-band Markovian modelling method we developed to detect LSB galaxies on images are described in Section~\ref{sec:markov}, the method to identify LSB galaxies from among the detected sources is presented in Section~\ref{sec:identification}, the results of tests to detect LSB galaxies on simulated images are given in Section~\ref{sec:simulations}, the results of the application of our method to two-band INT and three-band NGVS images are presented in Sections~\ref{sec:application} and \ref{sec:NGVS}, and the conclusions are presented in Section~\ref{sec:conclusions}.

\section{Markovian modeling \label{sec:markov}}

Our aim was to detect low surface brightness galaxy candidates, even with complex morphologies, through a simultaneous analysis of all available spectral bands. This multi-band inspection consists of modeling the observations with hidden (i.e., unobservable) variables, called \emph{labels}, which are spatially connected through a predefined neighborhood system. These labels correspond to discrete classes of objects with similar surface brightness. The main principle is to reconstruct the image of labels $X$ (also referred to as a segmentation map in the following) according to multiband observations $Y$. The Markovian modeling is very efficient in regions with low S/N (Salzenstein \& Collet 2006). It is therefore ideally suited to extract LSB galaxies from the noise.

Our method uses the principles of a Markovian approach on $X$, which are based on neighborhood relationship and the Bayesian inference linking statistically the multiband observations $Y$ to the segmentation map $X$. Since the end of the last century this methodology has provided us with a robust and efficient way to solve problems of restoration, detection, segmentation or classification. The power of such statistical image analysis lies in its specific regularization based on neighborhood behavior (Markovian modeling), coupled with an explicit noise model within each observation.

\subsection{MARSIAA}

We employ the dedicated MARSIAA (MARkovian Software for Image Analysis in Astronomy) modelling software for the segmentation of multiband images based on a hierarchical in-scale neighborhood system to overcome strongly noised observations (Laferte et al. 2000, Provost et al. 2004). In the following we give a brief summary of the method which is described in detail in the Appendix.

\begin{figure*} %1
 \plotone{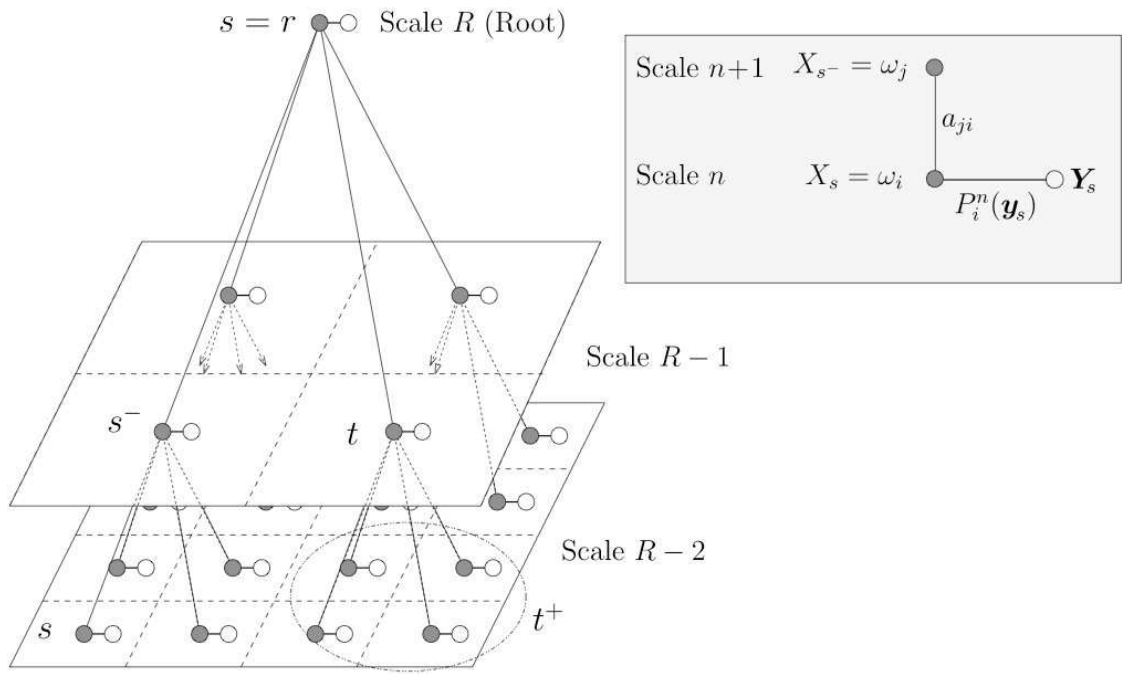}
 \caption{Dependency graph, corresponding to a quadtree structure.
 Filled black circles represent labels (segmentation classes $X$) and white circles represent observations (image pixels $Y$ at the highest resolution).
 At the bottom of the quadtree each label is linked to the corresponding pixel of the observations $Y$.
 Whereas the information contained in the observations is propagated upward within the quadtree, the labels are propagated downward according to 
 Markovian transmission probabilities, i.e. a label at a given level of the quadtree only depends on its parent. 
 } \label{fig:quadtree}
\end{figure*}

The segmentation principle, corresponding to a segmentation (labeling) process, is simple: the segmentation classes (labels) are described on a hierarchical quadtree (see Fig.~\ref{fig:quadtree}). At the bottom of the quadtree each label is linked to the corresponding pixel of the observations $Y$.
Whereas the information contained in the observations is propagated upward within the quadtree, the labels are propagated downward according to 
Markovian transmission probabilities, i.e., a label at a given level of the quadtree only depends on its parent. 
The quadtree structure is a well-adapted topology structure to express in-scale Markovian properties (transmission of information from parent to child), providing an efficient way to regularize the segmentation map (Laferte et al. 2000). This regularization maximizes the extent of objects within a
segmentation class even in the presence of strong noise.

Since the labels are unknown at the beginning of the process, they have to be estimated iteratively.
The quadtree structure deals with the label estimations at various spatial resolutions: the top represents the lowest resolution (a single label for the whole image) whereas the bottom corresponds to the finest resolution, i.e., the image pixels.
At the bottom of the quadtree all labels $X$ are linked to the observations $Y$ through a data-driven term defined as the parametric probability density function (pdf), which is assumed to be a multidimensional Gaussian of variable variance. 
The data-driven term allows the correct connection between the observations (scalar or multi-component at each location, according to the available data) and the label which has to be estimated at the same image location. The rules on labels between scales are managed by the Markovian property: the children labels depend on the parent label according to transition probabilities. The label at the finest resolution represents the end product of our algorithm: true or false for a detection problem, reconstructed surface brightness in a restoration problem, or $1,2,...,N$ for our segmentation problem where we sort image pixels into $N$ subsets of similar behavior. In our case the final map for a given class is a binary mask at the highest resolution.
This type of regularization scheme is generic, allowing the detection, segmentation, and classification of raw data:
\begin{itemize}
\item 
in the case of multiband observations, containing data at different scales (which is not our case), the quadtree is fed by the observations at different scales and the segmentation process takes into account all data simultaneously;
\item 
in the case of missing data, the quadtree can take into account only its prior (parent-children transition probability) to define the label.
\end{itemize}
Compared to other detection methods our approach is scale-free, because the hierarchical quadtree is only applied to the set of labels, whereas the information contained in the data at the highest resolution is not propagated to lower resolution planes within the quadtree. 
MARSIAA does not handle different PSFs explicitly. However, a band with data of significantly different resolution can feed a different level of the quadtree. 
On the other hand, the method will become scale-dependent if applied to data projected toward another representation space (wavelet, curvelet, shapelet, etc.) where the coefficients again may feed the quadtree at different levels (Crouse et al. 1998, Romberg et al. 2001, Murtagh et al. 2005, Starck \& Murtagh 2006, Chan et al. 2008), because structures are intuitively selected on dyadic scales. The advantage of our approach consists of the combination of multiband data in the Bayesian inference process, without any mathematical decomposition on a specific function basis: in this sense the data remain unchanged and the segmentation process is model-independent.

\subsection{MARSIAA objects \label{sec:marsiaaobjects}}

The segmentation process assigns image pixels to different classes. The final map of a given class is a binary mask in which all pixels belonging to this class have been assigned a value of 1 and where the rest are all 0. 
The mean amplitude of a given class depends on the information contained in the whole image and the number of classes.
Typically, class~0 is the noise and the LSB galaxies are contained in class~1 and/or 2. The user defines the total number of classes, which in our case was set to 6 for the INT images and 8 for the NGVS images, after tests using values between 4 and 10. Since a given pixel can only be assigned to one class, detected objects with a significant gradient in their surface brightness profiles will be distributed over different classes, e.g., the outer parts are labeled as class~1, the intermediate radii as class~2, and the center of the object as class~3. To obtain binary masks of each entire object in a given class $N_0$, we therefore have to add the binary masks of classes $N \ge N_0$.
The assignment of image pixels to the different classes depends on the total number of classes, the dynamic range of the image, and the surface brightness distribution. Our simulations (Sect.~\ref{sec:simulations}) show that the detection of known LSB galaxies on CCD images is hampered by the high dynamic range, due to the presence of stars. To avoid this problem we clipped the images at the 20$\sigma$ level ($\sigma$ being the noise in the images), which affects mainly bright stars but not the galaxies we are searching for. 
We determined the clipping level heuristically by applying our method to known LSB galaxies from Sabatini et al. (2003) which could not be detected in the image without clipping. The results are not sensitive to the exact value of the clipping level ($10$-$30 \sigma$).
The subsequent LSB detection was conducted on both the original and the 20$\sigma$-level clipped images for the INT data  and on the 30$\sigma$-level clipped images for the NGVS.

\subsection{The pipeline}

The MARSIAA segmentation map represents the heart of our proposed method for the detection of low surface brightness galaxies in
multiband images. The MARSIAA objects have then to be tested for stellarity and confusion. This is done with a subsequent algorithm called
DetectLSB which is based on radial surface brightness profiles to which an exponential profile is fitted. 
If a cut in exponential scale-length (e.g., $r_{\rm e} \ge 3''$) is
applied to the LSB galaxy candidates on images of intermediate depth ($\mu_{\rm B} \sim 26$~\masq), 
the resulting relatively small number of LSB galaxy candidates permits all to be visually inspected.
Since the number of LSB galaxy candidates is huge for deep images ($\mu_{\rm g} \sim 29$~\masq), we developped
additional selection software which is based on the source parameters from DetectLSB.
The whole processing pipeline consists of five steps: (i) cutting the images into subimages of sizes up to $2048 \times 2048$ pixels,
which represents the limiting image size for MARSIAA, (ii) application of MARSIAA simultaneously to the multiband subimages,
(iii) application of DetectLSB, (iv) cross-identification of the objects identified by DetectLSB via a clustering algorithm, and (v)
visual inspection of LSB candidate galaxies or application of a LSB selection software. The pipeline only needs human intervention between 
step (iv) and (v), to change the format of the source catalog from xml to ascii.

\section{LSB galaxy identification \label{sec:identification}}

Once the candidate objects are detected by MARSIAA, their surface photometry properties are examined and the objects are classified as potential LSB galaxies or other types, using a second software package, DetectLSB, which we developed specifically for this project.
The classification is based on the radial surface brightness profile, which should be exponential. Since LSB galaxies can have a dominant disk component, an elliptical fit to the isophotes is necessary.
DetectLSB is a fully automated C program designed to separate astronomical objects into two distinct classes: LSB galaxies and all others. It takes as input the original image, an associated segmentation map (from MARSIAA) and photometric information such as magnitude zero point and PSF. DetectLSB is applied to observations in a single band.
For each object identified in the segmentation map, it performs a two-dimensional ellipse fit and calculates the mean surface brightness profile over the radii of the estimated ellipse. It measures how well an exponential model fits the surface brightness profile, and finally, it provides a classification decision relying on deterministic filter rules. The LSB galaxies identified in different bands are then cross-identified via a clustering algorithm. Since different parts of the LSB galaxies can have different colors and artifacts are typically present in only one band, this ensures that all LSB candidates are robustly identified.

\subsection{Pre-processing}

Segmentation artefacts such as very small connected components or small holes are removed by using a morphological Open-Close-Close-Open (OCCO) filter\footnote{Morphological opening corresponds to an erosion operation followed by a dilation operation, whereas morphological closing corresponds to dilation followed by erosion.} (Gonzalez \& Woods 1992, Soille 2003) with a cross structuring element of 3 pixels width. A second filter eliminates objects having a diameter smaller than the PSF FWHM or larger than 200 pixels. Finally, objects too close to image borders (when the distance from center to border is less than 64 pixels) are removed and the remaining objects are labelled. 

\subsection{Ellipse fitting}

\begin{figure*} %2
 \plotone{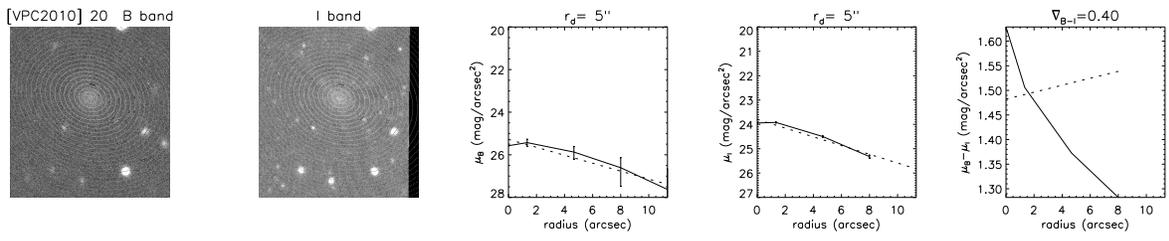}
 \caption{The LSB galaxy candidate [VPC2011]~20. From left to right: B and I band image, with the system of homothetic fitted ellipses superimposed; B and I band mean radial surface brightness profile; B-I color profile. The dashed lines show fitted exponentials to the radial profiles, and the resulting color gradient between the fitted profiles.} 
 \label{fig:exEllipse}
\end{figure*}

The isophotes of the galaxy are modelled as ellipses (Fig.~\ref{fig:exEllipse}), which are described by five parameters: center $c$, major axis radius $a$, minor axis radius $b$ and major axis position angle $\alpha$. The initial values are provided by Matlab through the function `regionprops' which determines the ellipse which has the same second moment as the object detected by MARSIAA. The fit is performed through the minimization of a cost function which decreases when symmetry increases. The space is divided into homothetic elliptical rings $\{e_a\}$ with a width of one PSF FWMH each. We define the partial cost function of each ring as the sum of the absolute difference between the mean value of pixels in each quadrants, whereas the total cost function is the weighted sum of all partial energies. Formally, the cost function $e$ is given by: 
\begin{equation}
e=\sum_{r=0}^{a}{G(r)\sum_{i=1}^{4}{\sum_{j=1,j\neq i}^{4}{|B_{i}(r)-B_{j}(r)|}}}
\end{equation}
where $B_{i}(r)$ is the mean pixel value over the $i$-th quadrant of the ring at radius $r$ and $G(r) = exp\left(\frac{-r^{2}}{2a}\right)$ is a weighting function which decreases with radius $r$.

The major and minor axis radii are defined as the last points on the radial fits whose error bar lies above the noise level.
The optimization is done with an adaptive gradient descent algorithm which processes the parameters sequentially, in the following order: center position, position angle and minor axis radius. The major axis, which is a direct result of MARSIAA, is not optimized. The minor axis radius and center parameters are constrained: the radius and the center have to stay within a range of ten pixels from their initial values, and the center within $\pm$ $10^{\circ}$ from its initial value. The procedure is run on a mean filtered version of the image ($3\times{}3$ pixel filter) and pixels belonging to other MARSIAA objects are ignored. If the estimated center is not included within the original MARSIAA object, we consider that the fit has failed and the object is eliminated.

\subsection{Surface brightness profiles and linear regression \label{sec:regression}}

In the next step, we compute the mean surface brightness profile using all 4 quadrants. An exponential profile is fitted by performing a linear regression on the mean surface brightness profile in a single band. Within each ellipse annulus $e_a$ we compute the mean surface brightness value $\mu_a$ and its standard deviation $\sigma_a$, using a mean filtered version of the image ($3\times{}3$ pixel filter), and ignore all pixels belonging to different MARSIAA objects and to segmentation classes $< j-1$ or $> j+1$, where $j$ is the class of the majority of pixels in the annulus. The local background level $\mu_b$ and local pixel noise $\sigma_b$ are estimated in a window of $512 \times{} 512$ pixels centered on the object with the same iterative clipping algorithm as used in SExtractor (Bertin \& Arnouts 1996). The linear regression is performed only on those points whose error bars lie above the background level and whose distance from the object center is smaller than that of the first point which lies below the background level (Fig.~\ref{fig:exEllipse}). We also calculate the half-light-radius and the magnitude based on the radial profile. For the determination of the half-light radius, the
surface brightness profile was not interpolated. The half-light radius has thus discrete values. We extrapolated the magnitudes by assuming an exponential
profile and a measured source extent of $3.4$ scale lengths. The extracted half-light radii and magnitudes are thus rough estimates and are not intended to
replace photometric measurements with more sophisticated methods.

\subsection{Filtering and clustering of the results}

We have defined a set of empirical selection rules based on estimated surface brightness profiles for the removal of spurious detection like stars, high surface brightness galaxies or multiple objects. An object is discarded if one of the following conditions is met:
\begin{itemize}
\item the linear regression includes less than 3 points;
\item multiple objects: the second point is higher than the first one and there exists another point higher than the second one (not necessary the third one); 
\item surface brightness criterion: the extrapolated central surface brightness is higher than $30\,\sigma$ in the B/g band and $60\,\sigma$ in  the I/i band;
\item star criterion: the difference between the extrapolated central surface brightness and the measured central surface brightness is more than $2$~\masq (this case applies when the wings of a star are fitted), where the latter is defined as the mean pixel value in a central ellipse with a major axis radius of 4 pixels;
\item extension criterion: the width of the extrapolated curve at $20\%$ of its height is less than $1.5$ times the width of the PSF at $20\%$ of its height. This criterion rarely applies.
\end{itemize}
Since the same object can be detected and identified as an LSB galaxy using masks of different classes, DetectLSB can yield up to 5 sets of derived parameters for the original and clipped image each (see Sect.~\ref{sec:marsiaaobjects}). We gather these sets of parameters to form one single object using a clustering algorithm based on the projected distances between the different identifications.

\section{LSB galaxies detection simulations \label{sec:simulations}}

\subsection{Method \label{sec:method}}

In order to evaluate the performance and the limits of our method, we performed several tests on simulated images. All simulations were done using single- and multi-band (2 and 3 bands) images. We considered two classes of sources -- elliptically shaped LSB disk galaxies with an exponential surface brightness profile, and point sources -- and a flat background. These images are convolved with the point spread function and Gaussian noise is added to them:
\begin{equation}
Y = H * m +b
\end{equation}
where $*$ stands for a convolution, $Y$ is the simulated image, $H$ the point spread function, $m$ the sum of all components of the simulation (LSB galaxy, point sources, background), and $b$ the noise (assumed to be Gaussian with zero mean).

We use the following image characteristics which are close to the INT observations to which we applied MARSIAA (see Sect. 5); the intensities are given in arbitrary units:
\begin{itemize}
\item pixel scale: \as{0}{33}/pixel (in all bands);
\item PSF FWHM: \as{2}{1} in the first band ($B$) and \as{2}{0} in the second ($I$);
\item background level $\mu$: $779.6$ in the first band and $3717.1$ in the second;
\item noise deviation $\sigma$: $16.9$ ($25.98$~\masq) in the B band and $31.6$ 
($24.82$~\masq) in the I band.
\end{itemize}

The free parameters of our simulations are: scale-length and central surface brightness of the LSB galaxy in both bands and the number of point sources. The size of the simulated images is $512\times{}512$ pixels. To avoid border effects, the LSB galaxy is always placed at the center of the image. It has a constant eccentricity of $\sqrt{3}/2$ (i.e., the major axis is two times longer than the minor axis) and a major axis position angle of $0$ degree. Point sources are generated randomly in an annulus with an inner radius of $75$ pixels and outer radius of $128$ pixels. Their peak brightness is also chosen randomly between $25\sigma$ and $1000\sigma$.

\subsection{Results \label{sec:simresults}}

\begin{figure*} %3
	\plotone{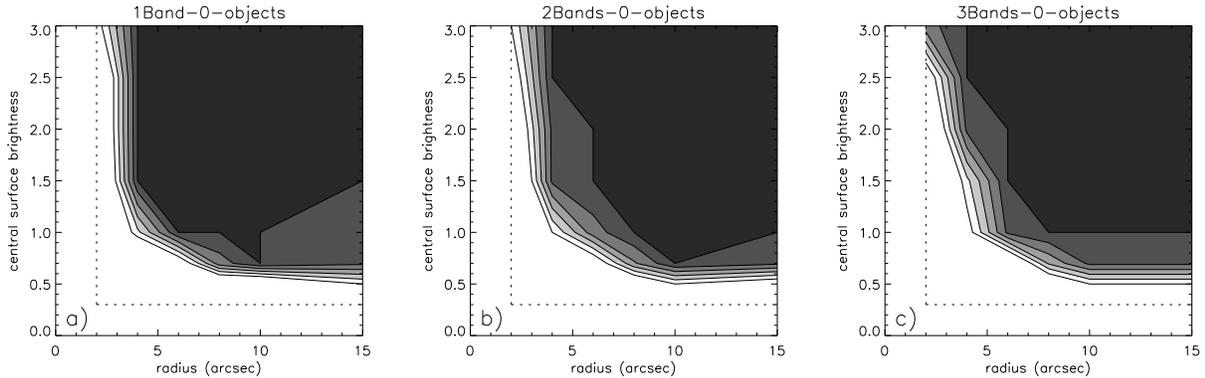} 
	\caption{Simulated LSB galaxies detection statistics as function of central surface brightness (in $\sigma$) and radius (in arcsec) for 1, 2 and 3 bands, respectively (from left to right). The FWHM of the PSF is $2''$. The simulated LSB galaxies have the same size and central surface brightness
in all bands. The contour levels are 0.5, 0.6, 0.7, 0.8, 0.9 and 1.0, from light to dark grey. The dotted lines indicate the borders of the parameter space. No polluting point sources were added. The detection statistics become worse for 2 and 3 bands; this is the well-known Hughes phenomenon.} 
	\label{fig:simulation0}
\end{figure*}

\begin{figure*} %4
	\plotone{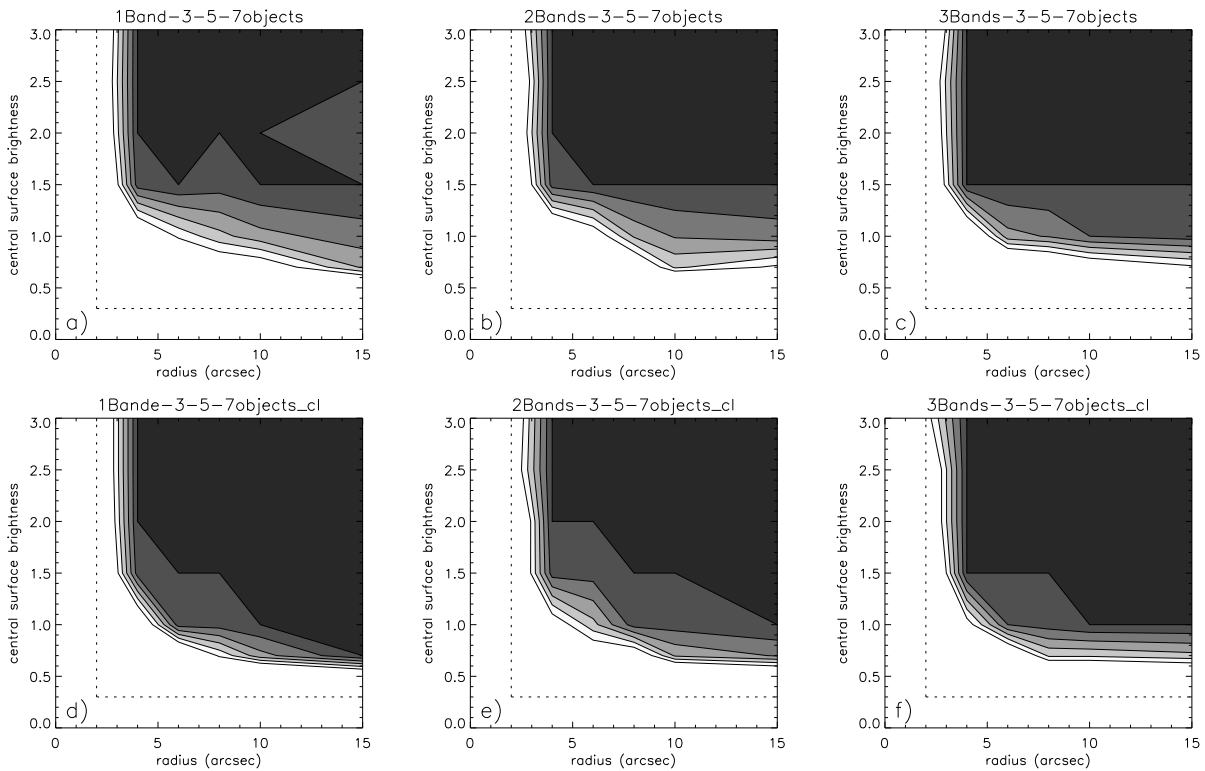}
	\caption{As in Fig.~\ref{fig:simulation0}, but in this case the images contained either 3, 5, or 7 polluting point sources. The FWHM of the PSF is $2''$.
	  The results of multiple simulations with different numbers of polluting sources where included in the detection statistics.
	Upper row: unclipped images. Lower row: images clipped at S/N=20.
	Compared to Fig.~\ref{fig:simulation0}, the presence of polluting point sources deteriorates the single-band detection statistics,
	but improves those of the multiple band cases.}
	\label{fig:simulation1}
\end{figure*}

We performed 10 sets of LSB detection simulations which are designed to reproduce INT data of Sabatini et al. (2003; see Sect.~\ref{sec:application}).
For each set an LSB galaxy with a central surface brightness between $0.3\sigma$ and $3\sigma$ and a characteristic scale-length between $2''$ and $15''$ was hidden in a $512 \times 512$ image. Three, five, and seven point sources were then added to the image in 21 different configurations. For each set, MARSIAA and DetectLSB were applied on a clipped and a non-clipped test image, because the division of the image pixels into the different classes depends on the dynamic range of the image and the surface brightness distribution (see Sect.~\ref{sec:marsiaaobjects}). Figs.~\ref{fig:simulation0} and \ref{fig:simulation1} show 
the detection statistics for the following six different scenarios:
\begin{enumerate}
\item
single band (a); 
\item
2 bands with LSB galaxies of the same central surface brightness and characteristic scale-length in both bands (b);
\item
3 bands with LSB galaxies of the same central surface brightness and characteristic scale-length in both bands (c);
\item
single band with a 20$\sigma$-level clipping (d); 
\item
2 bands with LSB galaxies of the same central surface brightness and characteristic scale-length in both bands, with 20$\sigma$-level clipping (e);
\item
3 bands with LSB galaxies of the same central surface brightness and characteristic scale-length in both bands, with 20$\sigma$-level clipping (f).
\end{enumerate}
While  in the simulations shown in  Fig.~\ref{fig:simulation0} no point sources were added, the simulated images of Fig.~\ref{fig:simulation1} contained either 3, 5, or 7 polluting point sources. The results of $21$ simulations with different numbers of polluting sources where included in the detection statistics of Fig.~\ref{fig:simulation1}. The detection fractions are upper limits of what can be expected for observations, because (i) the simulated galaxies are designed to
have exponential profiles and (ii) real LSB galaxies might be located near or overlap with other extended sources.

In the case of no polluting point sources the detection statistics becomes worse for 2 and 3 bands. This is the well-known curse of dimensionality (Hughes phenomenon): the number of pixels in a segmentation class (containing the LSB structure) may become relatively small and the segmentation process fails. Because MARSIAA is applied to subimages with sizes between $512 \times 512$ pixels and $1024 \times 1024$ pixels, this scenario is never met in astronomical images where a large number of polluting point sources is always present. 

As expected, the presence of polluting point sources makes single-band detection statistics worse (Fig.~\ref{fig:simulation1}). On the other hand, the addition of multiple bands improves the detection statistics in this case. Our method is able to detect with a 90\,\% success rate LSB galaxies with a characteristic scale-length of $\ge 4''$ down to the $1.5 \sigma$ central surface brightness level. 
For the unclipped images the detection rate increases only slightly for objects larger than $6''$ when MARSIAA is applied to 2 and 3 bands simultaneously. When applied to images which are clipped at a 20$\sigma$-level, the detection limit at the $90$\,\% success rate for 
extended objects ($\geq 8''$) decreases from $1.5\,\sigma$ to the $1\,\sigma$ level.
This illustrates the effect of the surface brightness distribution on the segmentation process and justifies the approach to apply our method on clipped images. We want to emphasize here that the test images are idealized cases and that the application of our method on actual INT observations (Sect.~\ref{sec:application}) has shown that our detection limit in practice lies between $1.5\sigma$ and $2\sigma$ in both the original and the clipped images.

\begin{figure*} %5
	\plotone{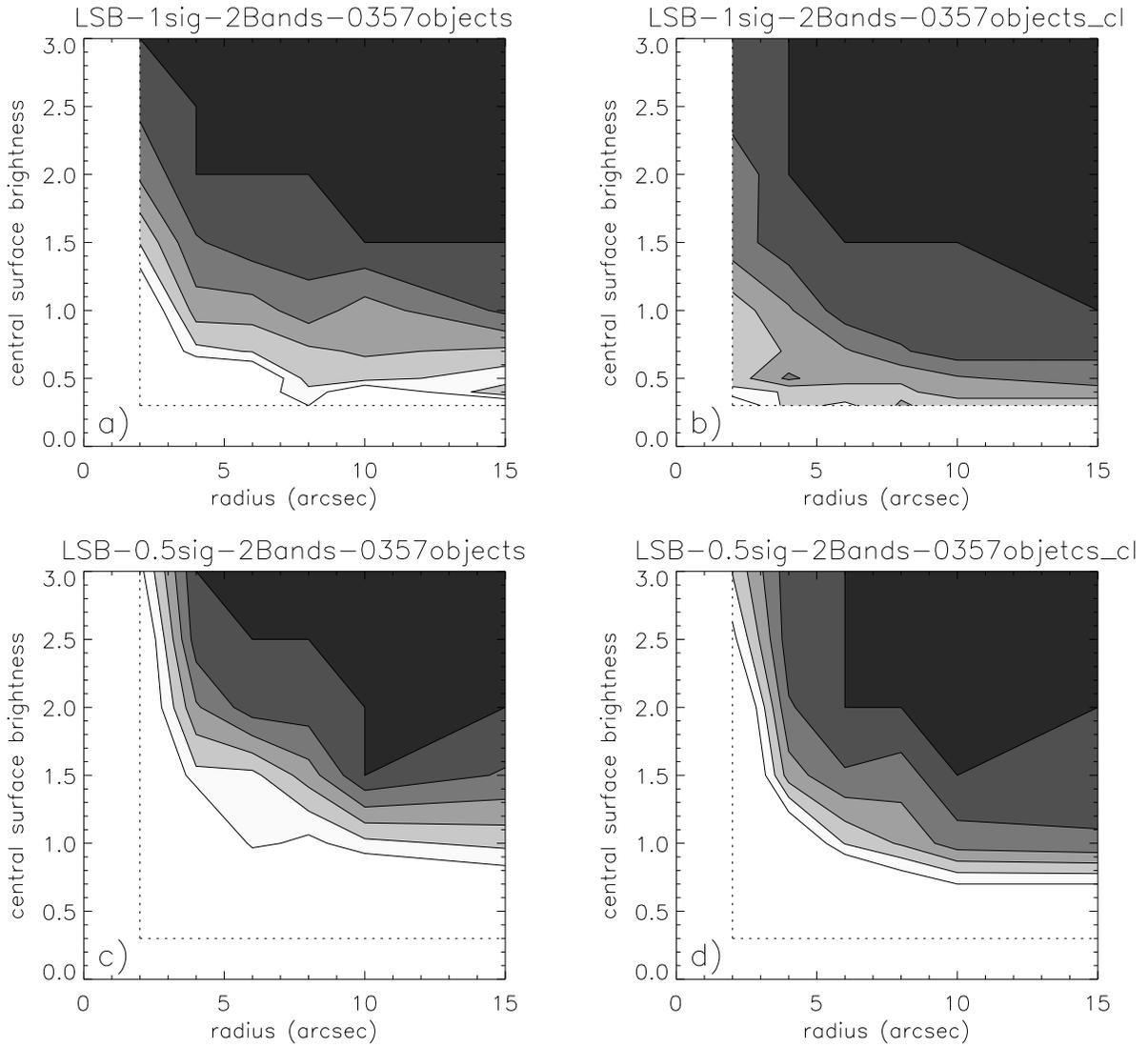}
	\caption{As in Fig.~\ref{fig:simulation0}, but in this case the LSB galaxy added to the first band has a fixed size ($10''$), central surface brightness and axial ratio, whereas the LSB in the second band has variable properties. The contour levels are 0.5, 0.6, 0.7, 0.8, 0.9 and 1.0, from light to dark grey. The dotted lines indicate the borders of the parameter space. Left column: unclipped images. Right column: images clipped at S/N=20. The addition of a second band improves the detection statistics.}
	\label{fig:simulation2}
\end{figure*}

A galaxy might have a different characteristic scale-length in each band. This can be the case for both late-type spirals, lenticulars, or ellipticals with a dominant old stellar population and only a small amount of recent star formation. To model such LSB galaxies of different colors, we have put an LSB galaxy of fixed size ($10''$), central surface brightness and axis ratio in one band and varied the properties of the second LSB galaxy in the second band (Fig.~\ref{fig:simulation2}):
\begin{enumerate}
\item
2 bands, with an LSB galaxy with a central surface brightness of $1\sigma$ in one band (a);
\item
2 bands, with an LSB galaxy with a central surface brightness of $1\sigma$ in one band, and 
 20$\sigma$-level clipping (b);
\item
2 bands, with an LSB galaxy with a central surface brightness of $0.5\sigma$ in one band (c);
\item
2 bands, with an LSB galaxy with a central surface brightness of $0.5\sigma$ in one band, and 
 20$\sigma$-level clipping (d).
\end{enumerate}

For LSB galaxies with different colors, i.e., different central surface brightness and/or different scale-length in the two bands, an object with a large scale-length ($10''$) and a central surface brightness of $1\,\sigma$ in one band can be detected with $90$\,\% probability if the surface brightness and scale-length in the second band are greater than $1.5\,\sigma$ and $4''$, respectively. For the clipped image the limiting surface brightness decreases below $1\,\sigma$ for LSB galaxies with a scale-length $> 6''$.
In an extreme case, where a large LSB galaxy ($10''$) has a central surface brightness of only $0.5\,\sigma$, its scale-length has to be larger than $9''$ in the second band to be detected at the $90$\,\% level with a central surface brightness of $1.5\,\sigma$. In this case, clipping does not significantly enhance the detection rate.

In conclusion, we are confident that in images with a PSF FWHM of about 2$''$ we can detect LSB galaxies with a radius $\ge 4''$ down to a central peak S/N$\sim$1.5. Large LSB galaxy (scale-length $\sim 10''$) located in one band, can in principle be detected down to S/N$\sim 1$. Clipping of the images always increases the detection probability.

\section{Application to INT B and I band images of Virgo cluster galaxies \label{sec:application}}

\subsection{INT observations}

The optical images we use in this paper are part of the 2.5-m Isaac Newton Telescope Wide Field Camera (INT WFC) survey of the Virgo cluster which was used by Sabatini et al. (2003, 2005) in their search for LSB dwarf galaxies in the cluster. Here we present results from the west-east B and I band strip in the cluster. The data were preprocessed and reduced using the Wide-Field Survey pipeline, see Sabatini et al. for details -- the photometric zero points are accurate to 1-2\,\%. The results are given in the B and I magnitudes of the Johnson-Cousin photometric system. Each $2048 \times 4100$ pixel INT image has a size of \am{11}{26}$\times$\am{22}{55}, a pixel size of \as{0}{33}$\times$\as{0}{33}, a mean PSF FWHM of \as{2}{1} and \as{2}{0} in the B and I band, respectively, and the B band sky noise level corresponds to $\sim$26 \masq.

\subsection{Results}

After the source identification by DetectLSB, about $\sim 40$-$50$ objects were found per INT image, giving in total $800$ objects
on $18$ INT images. However, this total includes not only genuine Virgo LSB galaxies, but also background galaxies and confused sources.
An inspection by eye represents the most secure way to separate these different classes of sources.
To facilitate the checking of the DetecLSB identifications by eye, we developed the java tool LSBExplorer which displays the object together with its measured parameters, surface brightness profiles, image cut-outs and segmentation maps. This readily permits the user to flag a source as an LSB galaxy candidate. At the end of the inspection the selected sources can be saved in xml or VOtable format. Despite the automatic rejection criteria (Sect.~\ref{sec:method}), there are still many spurious objects. As a first test, three of the authors (SS., WvD., BV.) independently sorted the sources in one INT image (v234c4) by eye and rejected the spurious sources. In this way, respectively 70\%, 60\% and 52\% of the sources were rejected by the three examiners; 47\% were rejected by all three, 13\% were rejected by two, and 12\% were rejected by only one.
Thus more than half of the sources are spurious. We did not attempt to make the automatic rejection criteria more stringent, because this might lead to uncontrolled rejections of relevant sources.

To investigate the nature of the spurious sources, the same three authors then sorted the sources in 10 INT images into different categories (Table~\ref{tab:rejection}). 
The overall rejection rate is 50-70\%, consistent with the aforementioned statistics of the simultaneous inspection of one INT image. Of these rejections 50-70\% are classified as multiple sources which are sometimes confused and 20-40\% could not be distinguished from extended low-level variations in the sky noise. After the inspection about 15-20 sources were left per INT image. Most of them are cosmologically dimmed background galaxies which are typically compact and elongated, and have  characteristic scale-lengths smaller than $3''$ (Sabatini et al. 2003).

\subsubsection{Virgo cluster LSB galaxies}

LSB galaxies located in the Virgo cluster at a distance of $\sim 17$~Mpc have characteristic scale-lengths larger than or equal to $3''$ (Sabatini et al. 2003). We used this criterion to reduce the number of sources that have to be inspected by eye. LSBExplorer has an option to display only sources with a characteristic scale-length larger than a given value. This decreases the number of sources to be inspected by eye from about 40-50 to about 10-20 sources per INT image.

\subsection{Comparison to detections with matched filters and SExtractor \label{sec:sextractor}}

\begin{figure*}
 \plotone{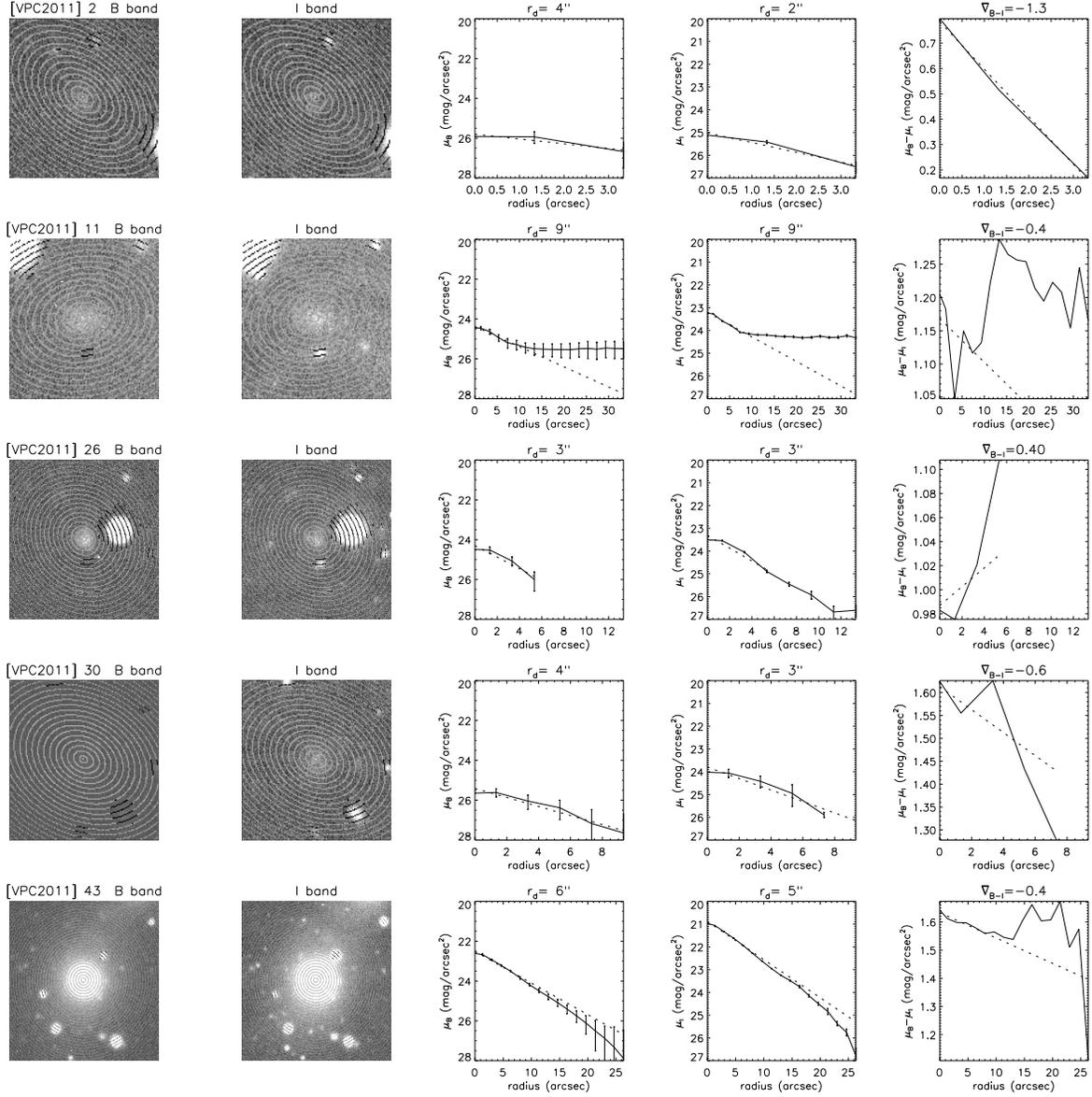}
 \caption{Selected LSB galaxy candidates with a characteristic scale-length larger than 3$''$ which were detected by us, but not by Sabatini et al. (2003) on the INT images.
  From left to right: B and I band images, with the system of homothetic fitted ellipses superimposed; B and I band mean radial surface brightness profiles; B-I color profile. The dashed lines show fitted exponentials to the radial profiles, and the resulting color gradient between the fitted profiles. 
 \label{fig:extract}}
\end{figure*}

We applied MARSIAA simultaneously to 18 INT B and I band images (1.3 square degree), and then applied DetectLSB separately to the B and I images using the MARSIAA segmentation masks (see Sect.~\ref{sec:marsiaaobjects}).
Table~\ref{tab:lsbtable} shows the 52 LSB galaxy candidates we found with a characteristic scale-length greater or equal than $3''$. Since the observed fields are overlapping, some sources appear twice in the table. We decided to leave these doubles, because they represent independent MARSIAA detections/DetectLSB identifications on images with different S/N. 
The table columns are: (1) galaxy number from this work (a star marks the 9 LSB galaxy candidates not found by Sabatini et al. 2003), (2) galaxy number from Sabatini et al. (2005), (3) right ascension, (4) declination, (5) central B band surface brightness, (6) B band scale-length, (7) I band characteristic scale-length, (8) I band central surface brightness, (9) color gradient $\nabla_{\rm B-I}=1/(\log(r_{\rm d}^{\rm B})-\log(r_{\rm d}^{\rm I}))$, (10) total apparent B magnitude, (11) total apparent I magnitude, (12) B-I color, (13) background flag. Based on the galaxy morphology we labeled 18 objects as background galaxies (bg in column 12). Among the remaining 34 objects we did not classify as background, we detected 9 that are not included in the Sabatini et al. (2005) catalog. Of these objects four ([VPC2011] 2, 11, 30, 43) have characteristic scale-lengths in the B band greater than $3''$ and three ([VPC2011] 1, 11, 43) have characteristic scale-lengths in the I band greater than $3''$.
[VPC2011] 1, 11, and 30 have other sources close to them (at $6''$-$10''$ distance) which could be removed by DetectLSB from their surface brightness profile, because they were identified by MARSIAA (see Fig.~\ref{fig:extract}). MARSIAA and DetectLSB are thus able to identify LSB galaxy candidates even in crowded regions of the sky. The recovered Virgo LSB galaxies are dwarf ellipticals and dwarf irregulars.

In Table~\ref{tab:comparison} we compare our detections of LSB galaxies with characteristic scale-lengths $\ge 3''$ with those of Sabatini et al. (2005) based on matched filters. In addition, we used SExtractor (Bertin \& Arnouts 1996) in a non-standard way tuned for our purpose (cut at $1.2 \sigma$ and with at least 200 connected pixels). 

Out of 28 objects detected with the matched filter analysis, MARSIAA+DetectLSB detect 20. Of the 8 objects that were not identified as LSBs by our method, [SDV2005] 57 is too close to an image edge to be detected by our method, [SDV2005] 2, 39, and 45 are very faint and not recognizable by eye as LSB galaxies on the images, and [SDV2005] 3, 38, 42, and 52 are most probably confused sources. Three objects were found by MARSIAA+DetectLSB but not by SExtractor: [SDV2005] 0, 1, and 35. On the other hand, two sources ([SDV2005] 3 and 41) were found by SExtractor and not by DetectLSB (although MARSIAA detected these, they were qualified as non-LSB by DetectLSB). 

We conclude that we detected all 20 LSB objects on the 18 INT images included in Sabatini et al. (2005) which we could classify by eye as bona fide LSB galaxies, and that we detected 3 objects that SExtractor, even in an optimized configuration, was not able to find.
Furthermore, we detected 4 new LSB galaxy candidates ([VPC2011] 2, 11, 30, 43) with characteristic scale-lengths $>3''$ which are not included in the Sabatini et al. (2005) catalog. Two of these candidates are very LSB $\mu_{\rm B} \sim 25$~\masq ([VPC2011] 2 and 30) and the others show pollution by nearby sources.

\begin{figure}
 \epsscale{0.7}
 \plotone{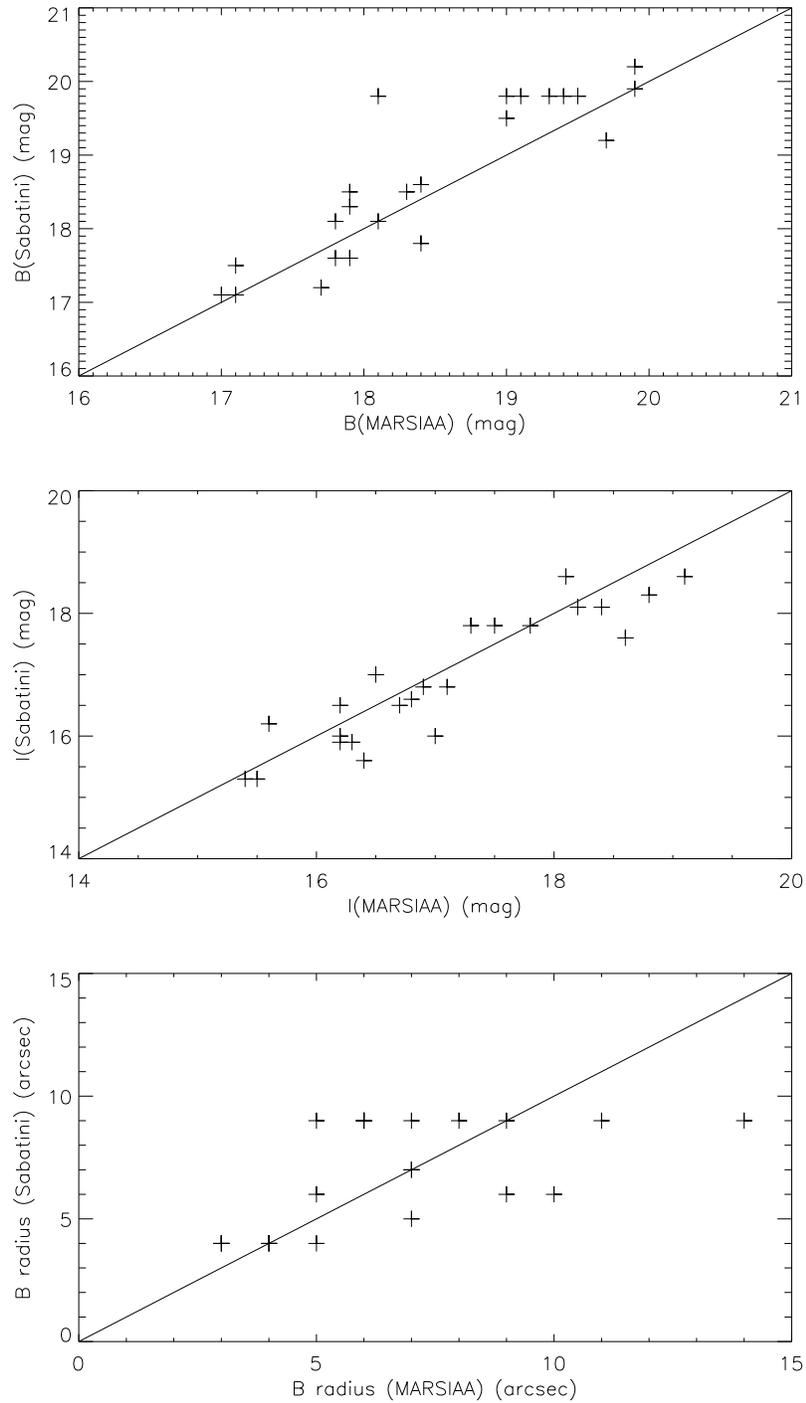}
 \caption{Comparison between  results of MARSIAA/DetectLSB and   Sabatini et al. (2003). Upper panel: total B magnitude; middle panel: total I magnitude; lower panel: characteristic scale-length on the INT iamges in the B band. The solid lines correspond to equality between both results.}
 \label{fig:lsbplots}
\end{figure}

In Fig.~\ref{fig:lsbplots} we compare our total B and I band magnitudes and B band characteristic scale-lengths with those of Sabatini et al. (2005). The correlation has a scatter of 0.5~mag in magnitude, and \as{2}{3} in radius. Given the approximate determination of our magnitudes based on simple exponential profiles, the large scatter is expected.

\section{Application to NGVS g,r, and i images \label{sec:NGVS}}

\subsection{NGVS observations}

The Next Generation Virgo Cluster Survey (NGVS)  uses the  MegaCam instrument on the 3.8m Canada-France-Hawaii Telescope (CFHT) to carry out a comprehensive optical imaging survey of the Virgo cluster, from its core to its virial radius covering a total area of 104 deg$^{2}$ in the u$^{*}$griz bandpasses
(Ferrarese et al. 2012). The results are given in CFHT magnitudes which are close to the SDSS magnitudes. 
All preprocessing of NGVS data is performed at the CFHT using the Elixir-LSB pipeline (see Ferrarese et al. 2012 for details), while the primary stacking for the NGVS is performed using a variant of the MegaPipe pipeline (Gwyn 2008). Each $\sim 20000 \times 20000$ pixel NGVS stacked image has a size of $1$~deg$^{2}$ with a pixel size of \as{0}{186}$\times$\as{0}{186}. The mean PSF FWHM is $\sim$\as{0}{8} in the g and r bands, and $\sim$\as{0}{6} in the i band.
The NGVS reaches a point-source depth of g$\sim 25.9$~mag (10$\sigma$) and a surface brightness limit of $\mu_{\rm g} \sim 29$~\masq (at $2\sigma$ above the mean sky level). The NGVS is thus $\sim 3$~mag deeper than the INT observations presented in Sec.~\ref{sec:application}.

Since the u$^{*}$ and z band NGVS data are at half-depth, LSB structures are generally detected at a lower significance level in these bands. Therefore, the detection statistics for the simultaneous application of MARSIAA to u$^{*}$griz images are worse than that for the application to the gri filters only.
As already stated in Sec.~\ref{sec:simresults}, in  empty bands the number of pixels in the  segmentation class that contains the LSB structures may become relatively small and the segmentation process fails (Hughes phenomenon). Therefore, we applied MARSIAA simultaneously to the  g, r, and i bands only, using  the following pipeline:
(i) each NGVS gri image is cut into $144$ subimages  of size $2048 \times 2048$ each, with  overlapping regions of $256$ pixels, (ii) MARSIAA is applied simultaneously to the gri subimages, 
(iii) DetectLSB, based on the MARSIAA segmentation maps, is applied sequentially to the NGVS g, r and i subimages, (iv) a clustering algorithm cross-identifies the sources detected by DetectLSB,
(v) selection software (see Sect.~\ref{sec:selection}) identifies the final LSB galaxy candidates.
The pipeline only needs human intervention between step (iv) and (v) to change the format of the source catalog from xml to ascii. We produced binary masks based on a clipping of the g band images at a surface brightness level of $26.7$~\masq ($\sim 30\sigma$). Higher and lower clipping values did not improve the results. After testing MARSIAA with $6$ to $10$ classes on a selected NGVS image from the Virgo cluster core (NGVS+0+1) in the gri bands, we set the number of MARSIAA classes to $8$.
The source clustering algorithm was run with cluster radii from $1''$ to $10''$. We adopted a cluster radius of $2''$,  which
gave the best results.

The pipeline was run  on a single processor, and required  $90$~h for MARSIAA, $90$~h for DetectLSB and $1$~h for the clustering and selection software for  a set of gri images of one  NGVS square degree  field. 
For comparison we also applied SExtractor with the following parameter sets: $2.5 \sigma$ and $5$ connected pixels,  and $1.5 \sigma$ and $200$ connected pixels. The resulting binary maps were added to yield a SExtractor segmentation map. DetectLSB and the selection software were then applied to these data as in the case of MARSIAA.
Obviously, SExtractor is significantly faster than MARSIAA:  SExtractor needs less than $1$~h of processor time for a set of gri images of one   square degree NGVS field,  and the application of DetectLSB to  the SExtractor maps needs $80$~h. Using the  pipeline with SExtractor on  a set of 3-band images  of one  square degree NGVS field  thus needs about $4$ days, or half of the  $8$ days needed with MARSIAA.

The advantage of MARSIAA lies in the increased depth of the segmentation maps compared to SExtractor. 
MARSIAA/DetectLSB allowed us to identify up to $40$\,\% more LSB galaxies than SExtractor/DetectLSB (Sect.~\ref{sec:ngvsmock} and \ref{sec:ngvs1}).

\subsection{LSB galaxy identification \label{sec:selection}}

As for the INT images, DetectLSB was applied to the g, r, and i images separately. The resulting catalogs were merged to yield a final source catalog.
In addition, an extra software searched for extended MARSIAA class 1 objects without higher classes.
The shallower  depth of the INT images made it possible to select LSB galaxy candidates from the final catalog based on the characteristic radius and to do the final identification by eye. 
This was not possible for the NGVS images where MARSIAA/DetectLSB finds $20,000$--$40,000$ LSB galaxy candidates per square degree. On the other hand, out of this large number only 138 objects were identified by eye as Virgo cluster LSBs by the NGVS team on one of the NGVS images (NGVS+0+1) near the center of the cluster -- only about $1$\,\% of the candidates in our catalog. Although our method is intended to detect and identify additional LSB galaxies, we have to select less than $\sim 10$\,\% of real Virgo LSB galaxies out of our LSB galaxy candidates. 
This is a very challenging task. To do so, we developed an additional software which is based on the MARSIAA segmentation maps and the results of DetectLSB for the different filter bands and segmentation classes. In particular, it takes advantage of the MARSIAA segmentation classes to separate objects in crowded fields and to exclude objects which are too small (half light radius smaller than $1''$), have unphysical colors or surface brightness profiles, are too compact (many MARISAA classes contained in the object), or highly asymmetric. This LSB identification led to about $1500$--$2000$ candidate LSB galaxies.
An inspection by eye of $400$ of these showed that the majority ($\sim 90$\,\%) are bona fide LSB galaxy candidates. Most of the candidates are not in the Virgo 
cluster, i.e., they are background galaxies. Spurious detections are mostly located in halos of extended sources (stars or galaxies). 
The majority of recovered Virgo LSB galaxies are dwarf irregulars, the recovered background galaxies are mostly spirals.
It thus turned out that we detect and identify elliptical (dEs on the INT images), disk (background), and irregular galaxies (dIs in the Virgo cluster).

\begin{figure*} %8
 \plotone{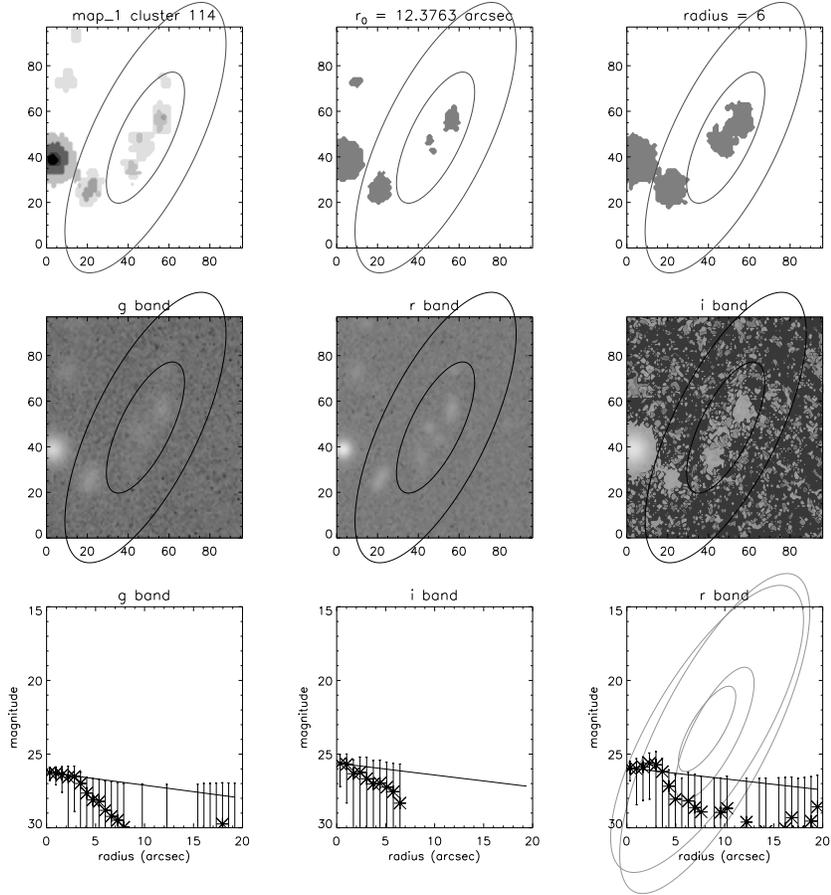}
 \caption{Example  of the application of MARSIAA/DetectLSB on NGVS gri-bandimages. Shown is a false positive result, due to confusion of three sources. Black ellipses: source extent (outer ellipse) and characteristic radius derived by DetectLSB (inner ellipse). Top row, from left to right: MARSIAA mask, SExtractor mask ($2.5\sigma$, $5$ connected pixels) and  SExtractor mask ($1.5\sigma$, $200$ connected pixels); middle row: g, r and i band images; lower row: g, r and i surface brightness profiles. Grey ellipses: source extents and characteristic radii derived by DetectLSB for the different segmentation classes and filter bands.
 \label{fig:crowding}}
\end{figure*}

\begin{figure*} %9
 \plotone{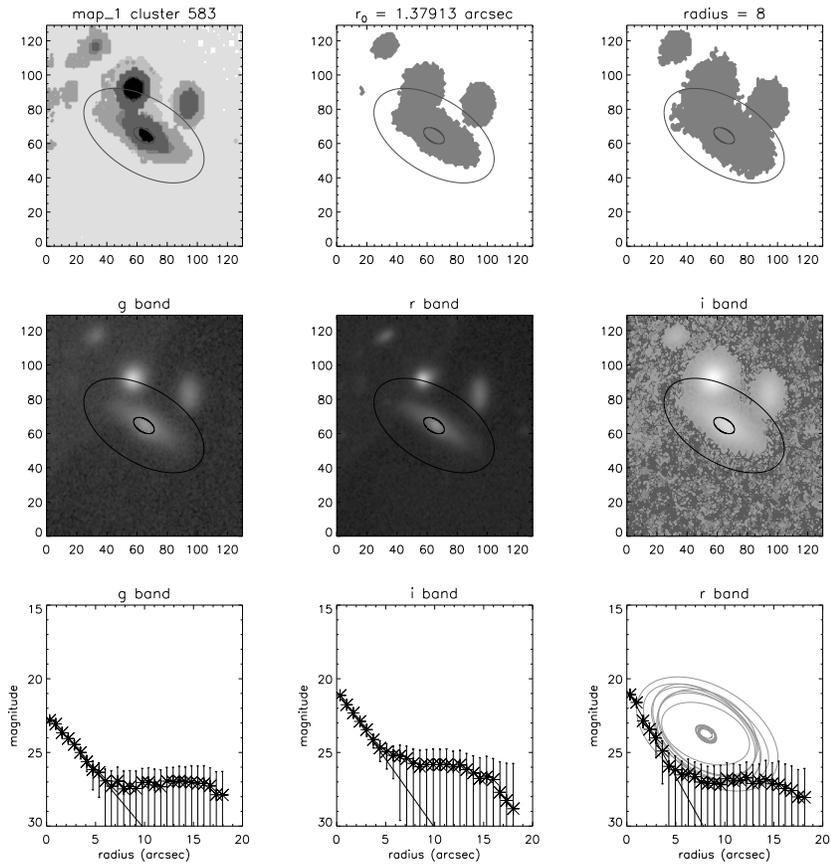}
 \caption{As in Fig.~\ref{fig:crowding}, but showing a confused, but detected and identified LSB galaxy in a crowded field with merging source halos
 \label{fig:crowding1}}
\end{figure*}

\begin{figure*} %10
 \plotone{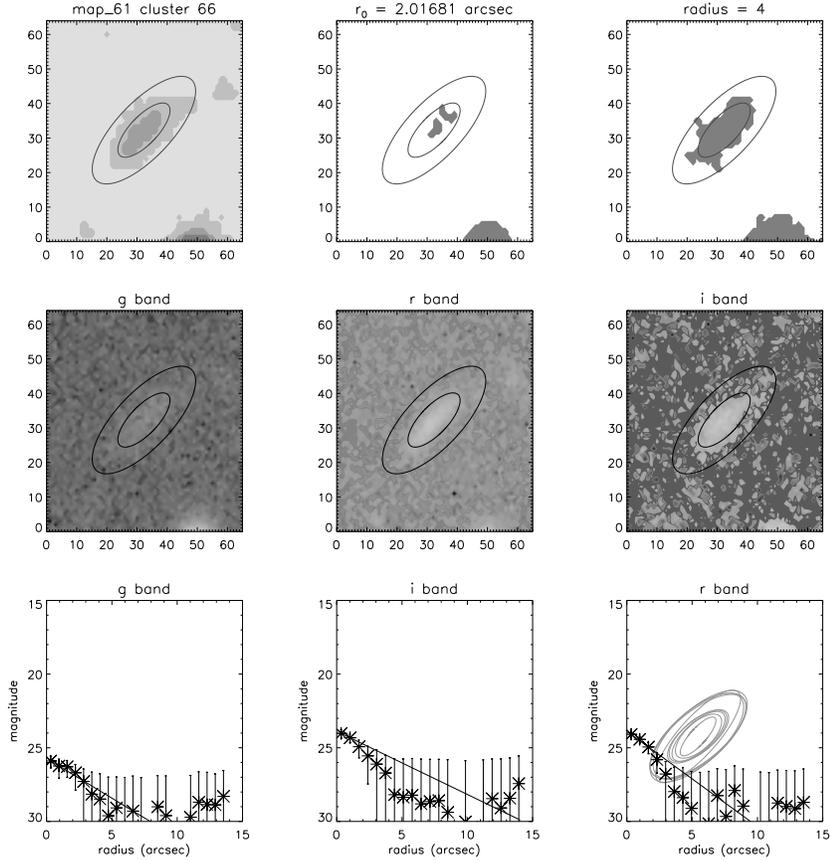}
 \caption{As in Fig.~\ref{fig:crowding}, but showing a typical detection and identification of an isolated LSB galaxy.
 \label{fig:isolated}}
\end{figure*}

At the faint surface brightness levels of the NGVS, source crowding and the merging of source halos become important problems. As an example, Fig.~\ref{fig:crowding} shows a MARSIAA/DetectLSB false LSB identification of multiple aligned sources.
Nevertheless, the LSB galaxy identification by DetectLSB is quite robust in crowded fields with merging source halos (Fig.~\ref{fig:crowding1}) since for the surface brightness profiles, different MARSIAA objects and pixels belonging to segmentation classes $< j-1$ and $> j+1$, where $j$ is the class of the majority of pixels in an ellipse annulus, are rejected . 
A typical detection and identification of an isolated LSB galaxy is presented in (Fig.~\ref{fig:isolated}).

In Figs.~\ref{fig:crowding}, \ref{fig:crowding1} and \ref{fig:isolated} we also show SExtractor masks for the two parameter sets ($2.5 \sigma$ and $5$ connected pixels,  and $1.5 \sigma$ and $200$ connected pixels). In general, MARSIAA produces deeper masks than SExtractor.
On the other hand, the increasing depth of the masks increases confusion. 
MARSIAA/DetectLSB detects and identifies about $25$\,\% more sources than SExtractor/DetectLSB. 

A secondary rejection based on color (r$-$i$ > 1.25-($g$-$r)), size and inclination (half-light radii smaller than $2''$ and inclinations $>45^{\circ}$),
and area + i band surface brightness lead to a total number of $\sim 1000$ LSB galaxy candidates per square degree NGVS image with only a small
decrease ($< 5$\,\%) in the identification statistics. Since the number of Virgo cluster LSB galaxies per NGVS square degree image is
about $100$ (Sect.~\ref{sec:ngvs1}), we are thus left with $\sim 90$\,\% of sources which are not Virgo cluster LSB galaxies.
Thus, about $1000$ have to be examined manually or by a smart
unsupervised fitting routine (e.g., GALFIT; Peng et al., 2010), which is beyond the scope of our work.

\subsection{Application to NGVS mock data \label{sec:ngvsmock}}

The NGVS team kindly provided us gri subimages of field NGVS-1+1 ($18000 \times 18000$ pixel) to which they added $396$ mock Virgo cluster LSB galaxies with g band magnitudes between 19 and 25. 
The galaxy luminosity  profiles are Sersic profiles (Sersic 1963) with variable effective radii ($0.6'' \leq r_{\rm e} \leq 5.6''$), exponents 
($0.6 \leq n \leq 2.0$) and axis ratios ($0.2 \leq b/a \leq 1.0$). 
Moreover, a nucleus was added to each mock object with $m_{\rm nuc}=m_{\rm gal} + 6.5$, $r_{\rm e}^{\rm nuc}=0.1 \times r_{\rm e}^{\rm gal}$ and $n=1$.
In addition to the mock Virgo galaxies, NGVS-1+1 also contains real LSB and HSB galaxies.
Both MARSIAA/DetectLSB and SExtractor/DetectLSB were applied to these images.
The two SExtractor binary masks ($2.5\sigma$ and $5$ connected pixels, and $1.5\sigma$ and $200$ connected pixels) were added to produce a segmentation map. DetectLSB was then applied to this map.

DetectLSB based on the MARSIAA segmentation map found $40,635$ objects separately in the g, r, and i band, the final source catalog (after cross-identification via a cluster algorithm) contains $11,417$ sources, and $263$ out of the $396$ mock Virgo LSB galaxies were detected.
DetectLSB based on the SExtractor segmentation map found $21,915$ separately in the g, r, and i band, the final source catalog contains $8721$ sources, and
$242$ out of $396$ mock Virgo LSB galaxies were detected.
We then applied the additional software described in Sect.~\ref{sec:selection} to the final source catalogs which reduced
the number of positive sources significantly:
with MARSIAA/DetectLSB we obtained $2089$ LSB galaxy candidates: $191$ out of the $396$ mock Virgo LSB galaxies were identified.
With SExtractor/DetectLSB we obtained $1606$ LSB galaxy candidates: $157$ out of $396$ mock Virgo LSB galaxies were identified.
As stated in Sect.~\ref{sec:selection}, the application of secondary rejection criteria based on color, size, and inclination
reduces the number of LSB galaxy candidates to $\sim 1000$ with a less than $5$\,\% decrease of the identification statistics.
A visual inspection of a portion of the LSB galaxy candidates not identified as mock galaxies showed that about $90$\,\%  
are real galaxies located in the background.

Thus, DetectLSB/MARSIAA detected and identified $\sim 30$\,\% more LSB sources than DetectLSB/SExtractor on the NGVS images ($11,417$ compared to $8721$).
Concerning the mock data, DetectLSB/MARSIAA detected $\sim 10$\,\% more mock Virgo LSB galaxies and identified $\sim 20$\,\%
more of them  than DetectLSB/SExtractor. It should be noted here that, since the mock Virgo LSB galaxies are relatively small, 
 confusion is less important and SExtractor gives results comparable to MARSIAA. LSB galaxies identified
by eye by the NGVS team can be much more extended than the mock objects and their identification based on SExtractor maps becomes more difficult
(see Sec.~\ref{sec:ngvs1}).

\begin{figure} %11
 \plottwo{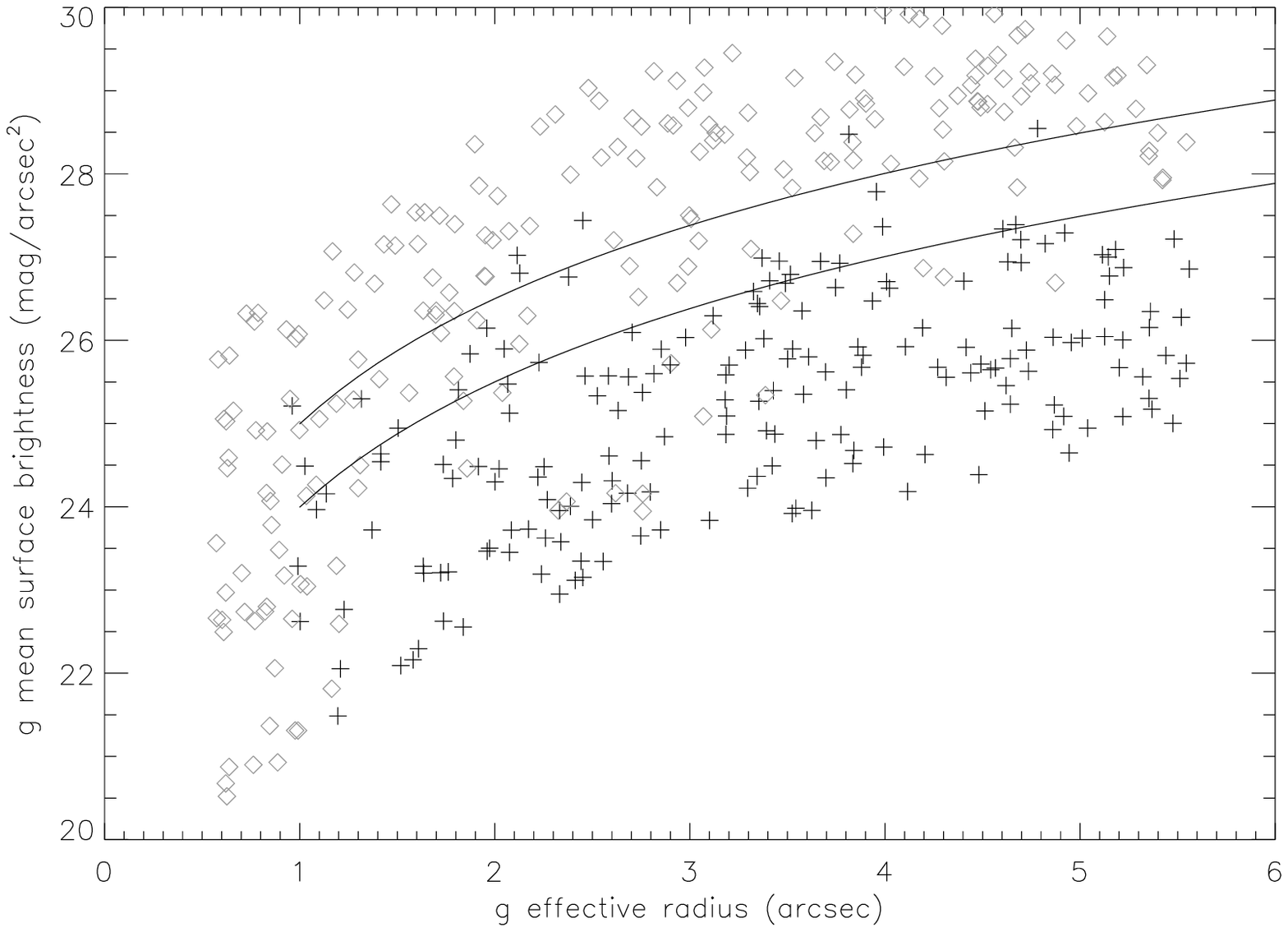}{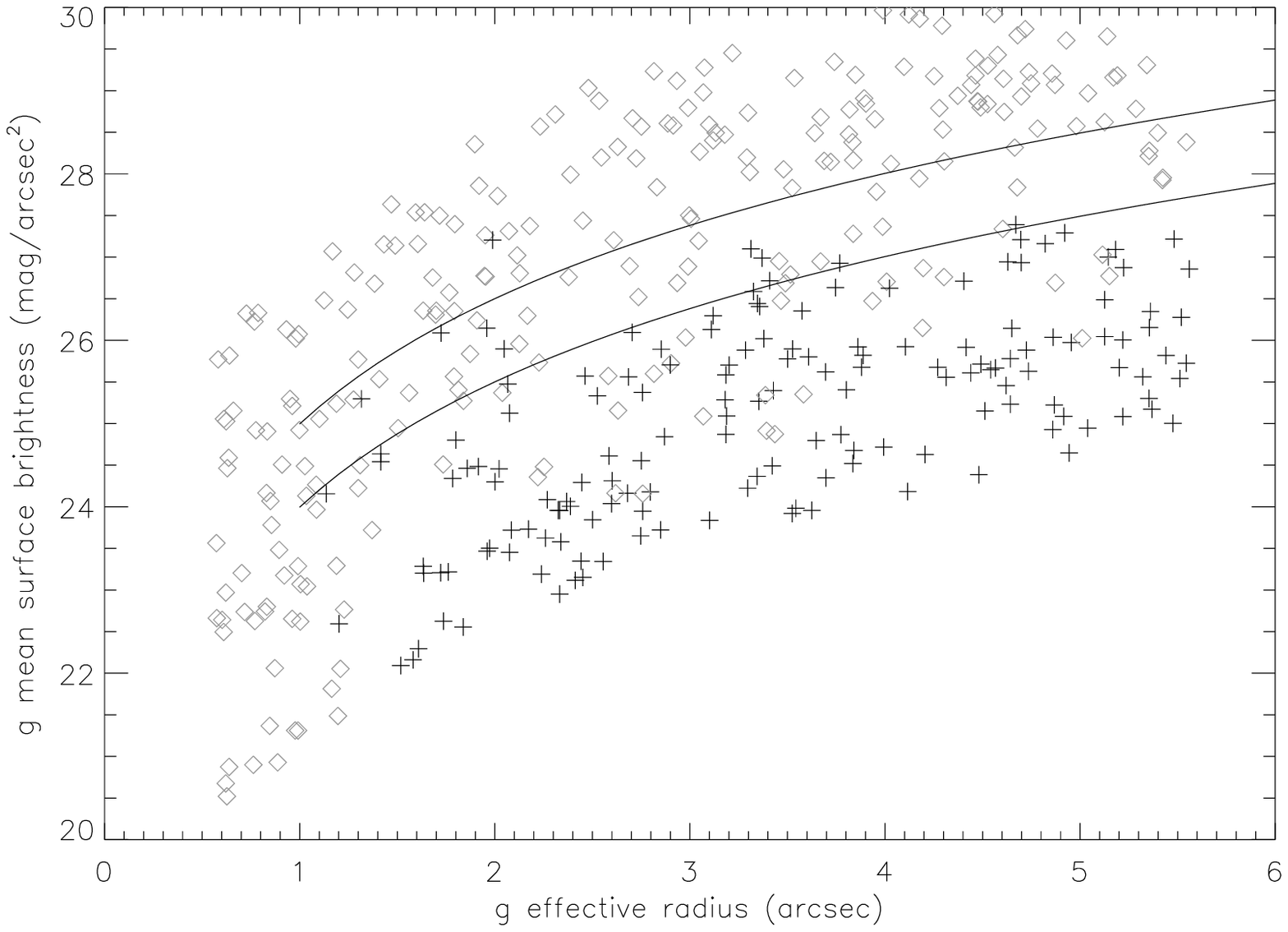}
 \put(-350,30){\bf MARSIAA}
 \put(-100,30){\bf SExtractor}
 \caption{Results for the mock data: mean g-band surface brightness as a function of  effective radius. Left panel: MARSIAA/DetectLSB. Right panel: SExtractor/DetectLSB. The solid lines correspond to g magnitudes of 22 and 23 (upper line). Grey diamonds: not identified mock Virgo LSB galaxies. Black crosses: mock Virgo LSB galaxies identified by MARSIAA/DetectLSB or SExtractor/DetectLSB. 
The completeness of both methods decreases significantly for $m_{\rm g} > 22$~mag, but MARSIAA/DetectLSB identified twice as many objects in this range as SExtractor/DetectLSB.
 \label{fig:modelgraph}}
\end{figure}

While SExtractor directly yields the magnitude and half-light radius for a given object, the application of DetectLSB results in several magnitudes 
and half-light radii depending on the fitted MARSIAA segmentation class. During the LSB galaxy identification, DetectLSB assigns a unique magnitude and half-light radius to each MARSIAA object.
The g mean surface brightness of the identified mock Virgo LSB galaxies as a function of the effective radius is presented in Fig.~\ref{fig:modelgraph}. The completeness for $r_{\rm e} > 1.5''$ and $m_{g} < 22$~mag is $90$\,\% for DetectLSB/MARSIAA and $83$\,\% for DetectLSB/SExtractor. The completeness of both methods decreases significantly for sources with $m_{g} > 22$~mag. DetectLSB/MARSIAA identified twice as many mock Virgo LSB galaxies in this magnitude range 
than DetectLSB/SExtractor. Whereas DetectLSB/MARSIAA is able to detect and identify LSB galaxy candidates up to a mean surface brightness of $m_{\rm g}=28.5$~mag (S/N$=1.6$; the NGVS surface brightness limit is $\sim 29$~\masq at $2\sigma$ above the mean sky level), confirming the results obtained from pure simulations (Sec.~\ref{sec:simresults}), a completeness of $90$\,\% is reached for sources with $r_{\rm e} > $\as{1}{5}  at $\mu_{\rm g}=26$~\masq and for sources with $r_{\rm e} > 3''$ at $\mu_{\rm g}=27.7$~\masq.

\begin{figure*} %12
 \plottwo{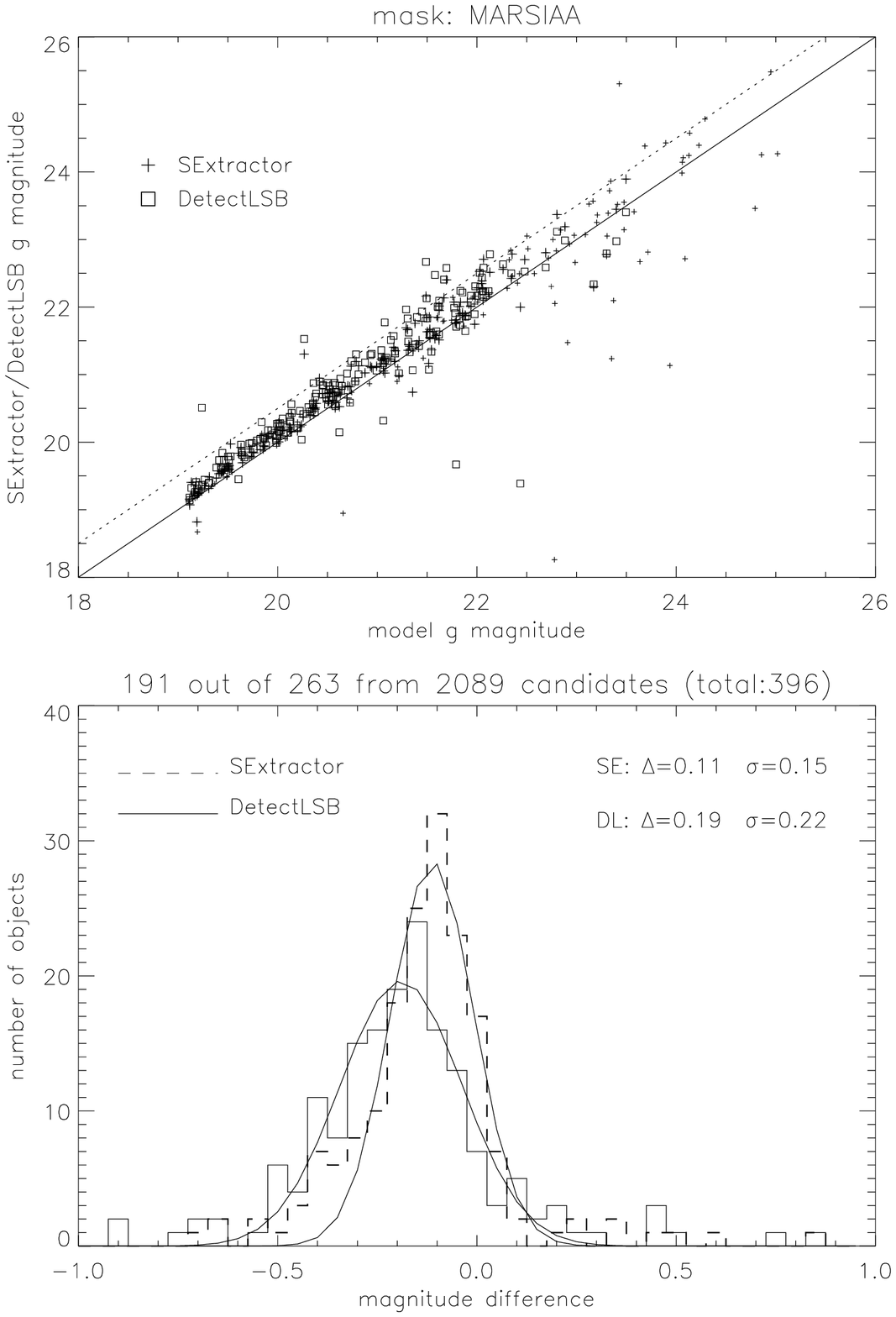}{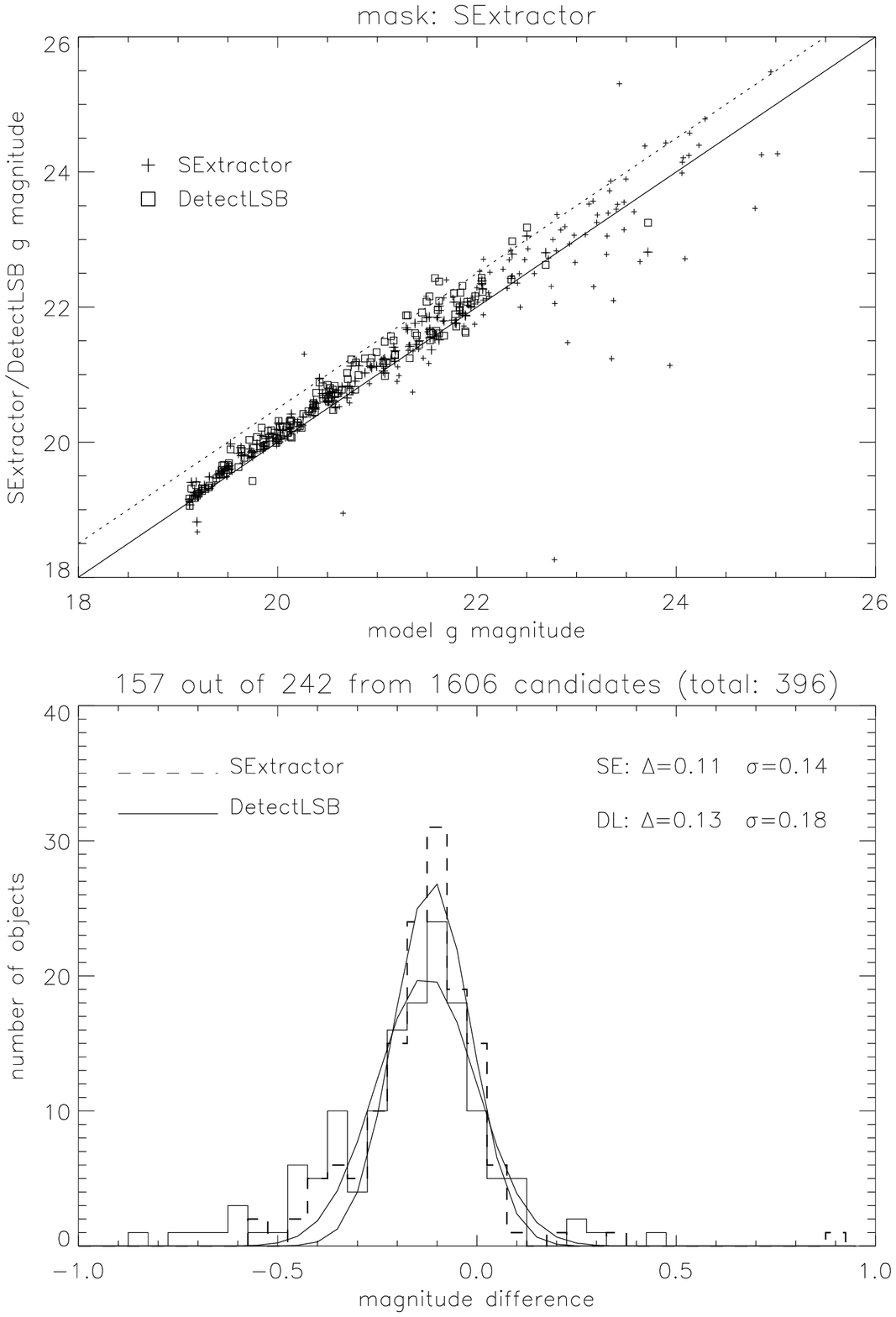}
 \caption{Results for the mock data: g-band magnitudes of the mock Virgo LSB galaxies recovered by SExtractor and by DetectLSB as a function of  their input g magnitudes. Left panels: MARSIAA/DetectLSB. Right panels: SExtractor/DetectLSB. 
Upper panels: small crosses: detected but not selected mock Virgo LSB galaxies; open squares: DetectLSB magnitudes and half-light radii. The solid line represents equality between the recovered and the input magnitude; to guide the eye, the dotted line has an offset of $+0.5$~mag. Lower panels: distribution of the difference between the recovered and the input g band magnitudes measured using SExtractor or DetectLSB, together with Gaussian fits. For the MARSIAA mask, using DetectLSB yields a larger dispersion and offset than SExtractor, whereas using both on the SExtractor mask results in comparable values.
 \label{fig:modelmag}}
\end{figure*}

We note that our magnitudes and sizes are rough estimates and are not intended to replace proper magnitudes and sizes derived from more sophisticated methods.
The recovered DetectLSB and SExtractor magnitudes are compared to the mock input magnitudes in Fig.~\ref{fig:modelmag}.
For the SExtractor magnitudes we use  preferentially those of the run with ($1.5\sigma$, $200$ connected pixels).
In the case of the application of DetectLSB based on SExtractor masks, the magnitude offsets and dispersions are comparable (offset: DetectLSB$=0.13$~mag, SExtractor$=0.11$~mag; dispersion: DetectLSB$=0.18$~mag, SExtractor$=0.14$~mag). This result validates the flux extraction
by DetectLSB compared to SExtractor.
In the case of the application of DetectLSB based on MARSIAA masks, the DetectLSB magnitudes show a larger  dispersion and a larger offset from the input magnitudes than the SExtractor  magnitudes (offset: DetectLSB$=0.19$~mag, SExtractor$=0.11$~mag; dispersion: DetectLSB$=0.22$~mag, SExtractor$=0.15$~mag). The DetectLSB magnitude offset and
dispersion is due to the quality of the unsupervised ellipse fit, which mainly depends on the first ellipse fit on the SExtractor or MARSIAA mask of the object. 
MARSIAA provides the deeper masks, SExtractor the cleaner masks. 
Since only pixels of similar segmentation classes are taken into account during the extraction of the surface brightness profile and
the magnitude is derived from this profile, the fluxes based on MARSIAA masks are lower than those based on SExtractor masks, despite
the higher depth of the MARSIAA masks.  

\begin{figure*} %13
 \plottwo{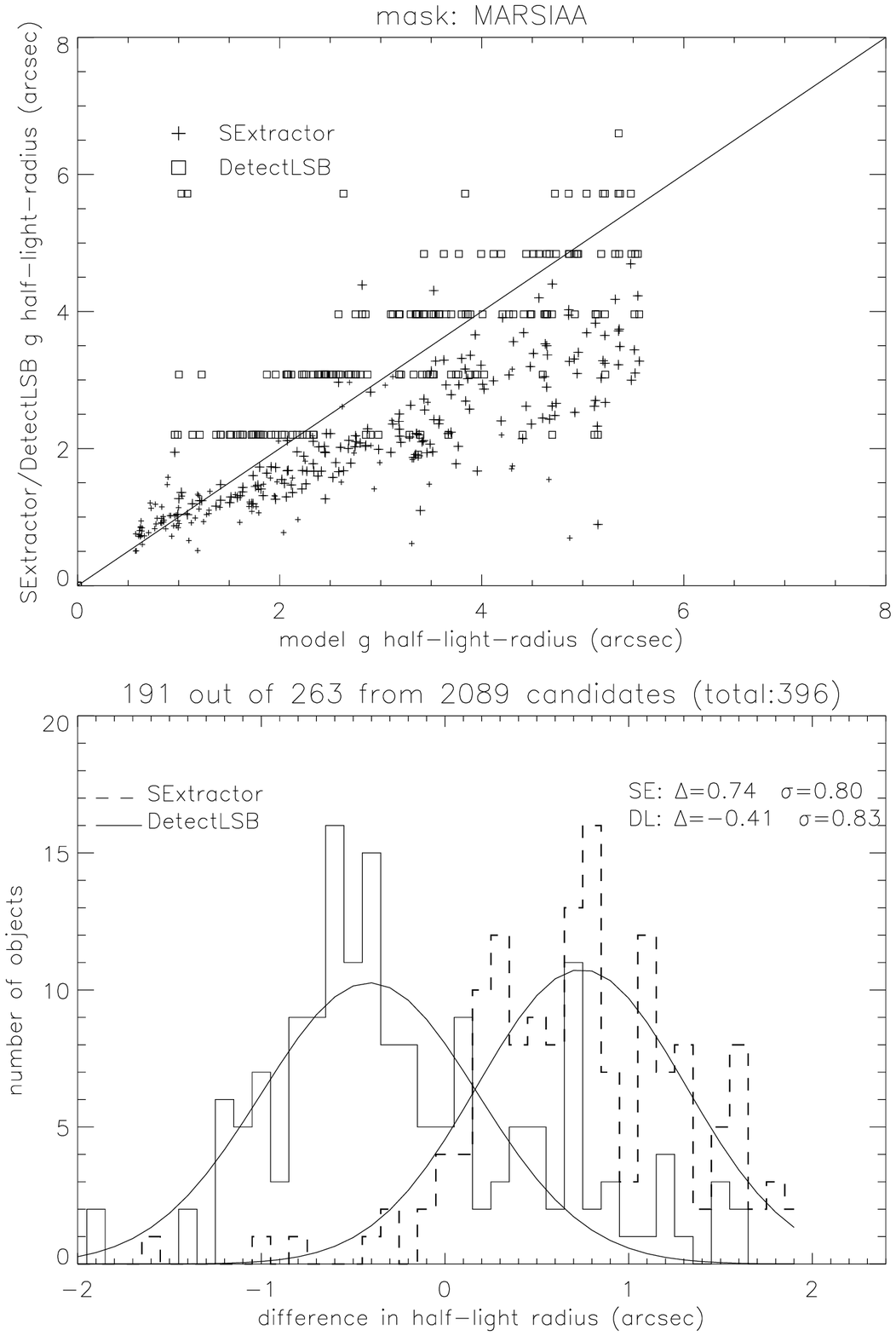}{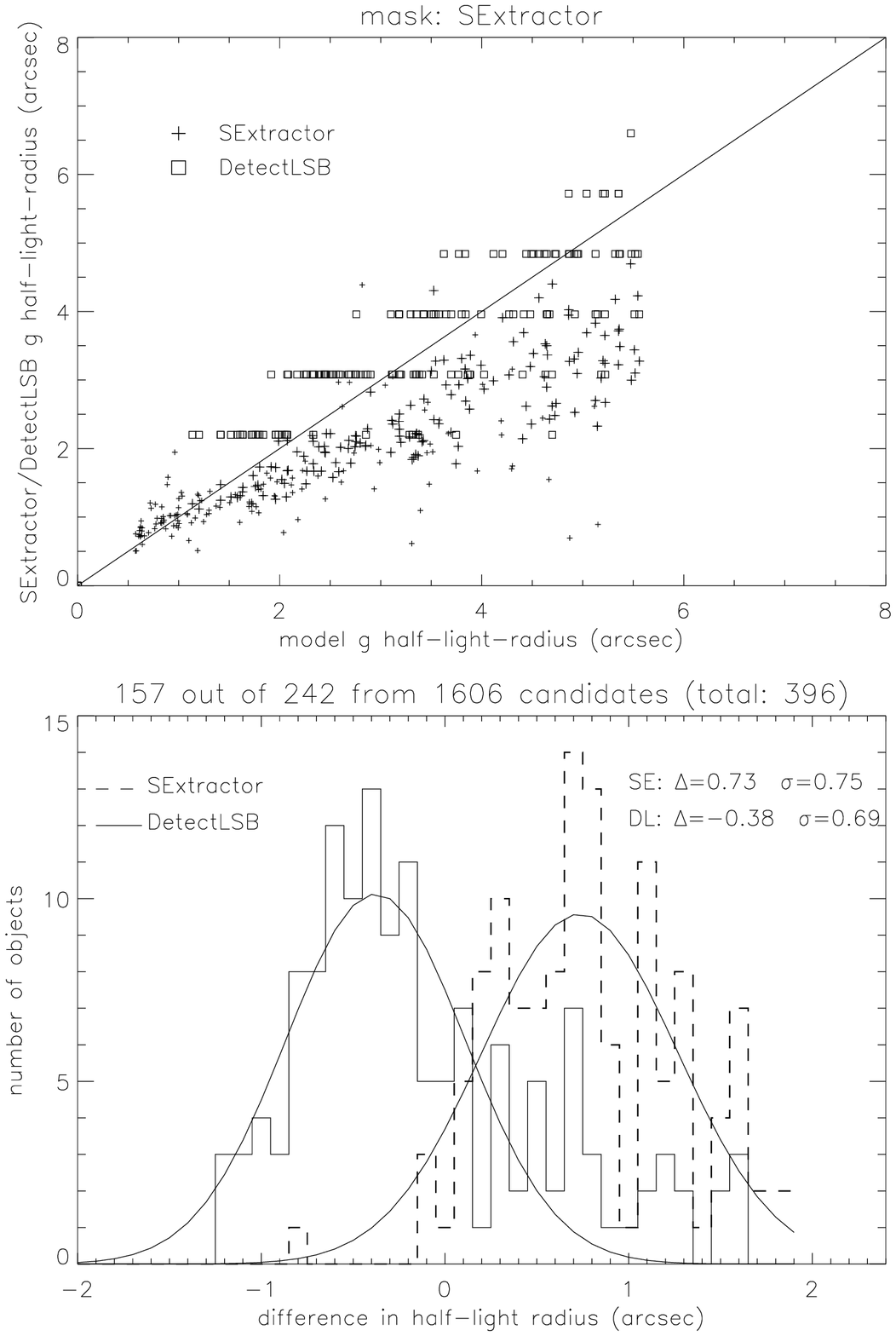}
 \caption{Results for the mock data: recovered SExtractor and DetectLSB effective radii of the mock Virgo LSB galaxies as a function of their input effective radii. Left panels: MARSIAA/DetectLSB. Right panels: SExtractor/DetectLSB. Upper panels: small crosses: detected but not selected mock Virgo LSB galaxies. The solid line represents equality between the recovered and the input effective radius. Lower panels: distribution of the difference between the recovered and the input effective radius together with Gaussian fits.
For radii $> 1.5''$, SExtractor underestimates the radii by $1''$, a value that increases with radius, whereas DetectLSB overestimates by an about constant offset of $0.5''$. 
 \label{fig:modelradius}}
\end{figure*}

The DetectLSB and SExtractor half-light radii are compared to the mock input half-light radii in Fig.~\ref{fig:modelradius}.
For input half-light radii exceeding  \as{1}{5} SExtractor underestimates the input radii on average by $1''$, and  the offset increases with increasing half-light radius.
On the other hand, DetectLSB overestimates the input half-light radii by an approximately constant offset of  \as{0}{5}. 
The dispersion of the differences between the input and estimated half-light radii is in all cases $\sim$  \as{0}{8}.
As stated in Sect.~\ref{sec:regression} the DetectLSB half-light radius is discrete and critically depends on the first ellipse fit on the SExtractor or MARSIAA 
mask of the object.

\subsection{Application to a square degree NGVS image \label{sec:ngvs1}}

As a further test, we applied MARSIAA/DetectLSB and SExtractor/DetectLSB to the NGVS image NGVS+0+1, the field east of NGVS-1+1.
On this image the NGVS team identified $138$ Virgo LSB members after visual inspection. This identification was based on surface brightness, extent, color, and morphology.
While the completeness of this by-eye catalog cannot be rigorously quantified, it is expected to be fairly complete. The average source extent of this sample is significantly larger than that of the Virgo mock galaxies, and  about $20$ of these LSB galaxies have half-light radii exceeding $10''$. 

\begin{figure*} %14
 \plotone{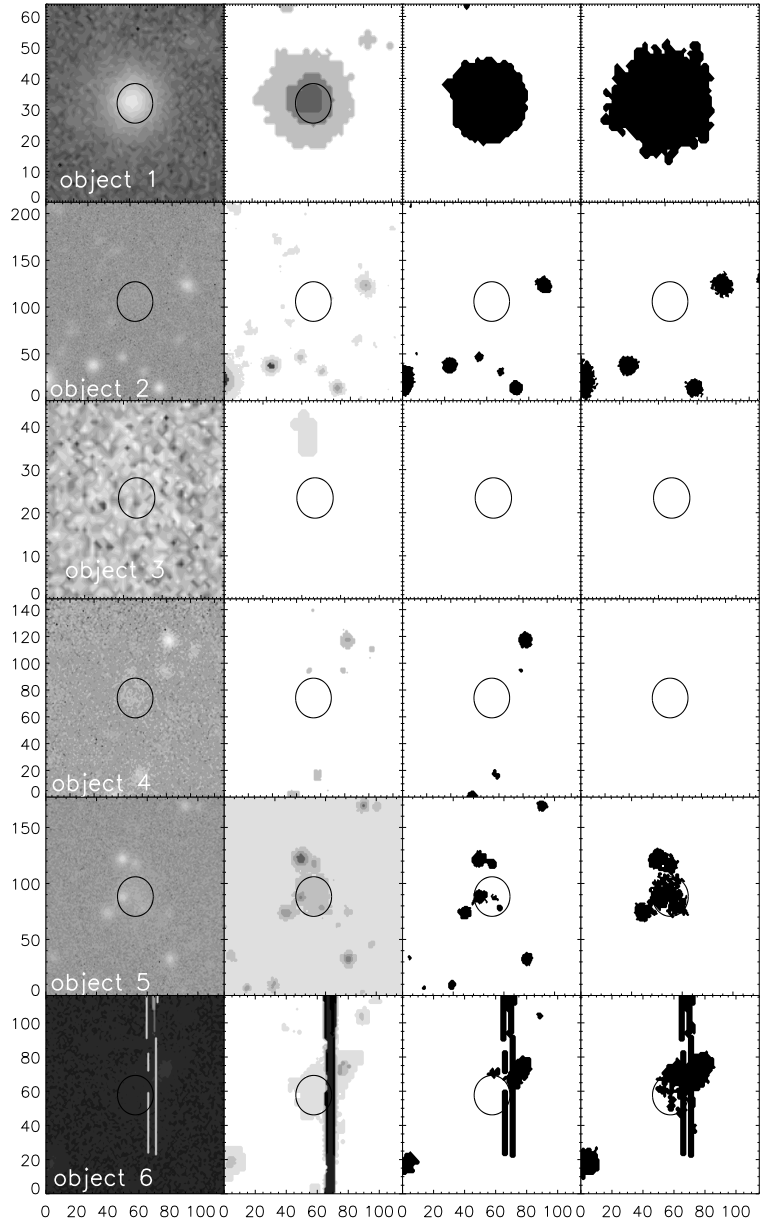}
 \caption{Examples of MARSIAA/DetectLSB pipeline for non-identifications of selected-by-eye LSB galaxies on NGVS+0+1: too small objects (object~1), too faint objects (object~2--4),
confused objects (object~5), image artifact (object~6). From left to right: g band image cutout, MARSIAA mask, SExtractor mask (classical parameters:
$2.5\sigma$ and 5 connected pixels), SExtractor mask (modified parameters: $1.5\sigma$ and 200 connected pixels). The $x$ and $y$ axis are pixels.
 \label{fig:ngvs_examples}}
\end{figure*}

We applied MARISAA, SExtractor, and DetectLSB with the same parameters as before.
DetectLSB based on the MARSIAA segmentation map found $125,787$ objects separately in the g, r, and i band, resulting in a final source catalog with  $33,119$ sources.
This catalogue contains $124$ out of $138$ Virgo LSB galaxies.
DetectLSB based on the SExtractor segmentation map found $76,797$ separately in the g, r, and i band, resulting in a  final source catalogue with  $25,913$ sources
containing $105$ out of $138$ Virgo LSB galaxies.
As for the mock data, we then applied the additional software described in Sect.~\ref{sec:selection} to the final source catalogs.
With MARSIAA/DetectLSB we obtained $2554$ LSB galaxy candidates: $115$ out of $138$ Virgo LSB galaxies were identified.
The non-detected LSB galaxies were confused, too faint (no object in the MARSIAA segmentation map), or too small (apparent size on the g band image
smaller than $2''$). In addition, one LSB galaxy is split into two parts by an image artifact (Fig.~\ref{fig:ngvs_examples}).
With SExtractor/DetectLSB we obtained $1813$ LSB galaxy candidates: $82$ out of $138$ Virgo LSB galaxies were identified.
Thus, DetectLSB/MARSIAA detected and identified $\sim 40$\,\% more LSB sources than DetectLSB/SExtractor on the two NGVS images.
As stated in Sect.~\ref{sec:selection}, the application of secondary rejection criteria based on color, size, and inclination
reduces the number of LSB galaxy candidates to $\sim 1000$ with a less than $5$\,\% decrease of the identification statistics.
As for the mock data, a visual inspection $\sim 400$ LSB galaxy candidates not identified as Virgo LSB galaxies showed that about $90$\,\% 
are real galaxies located in the background.

\begin{figure} %15
\plottwo{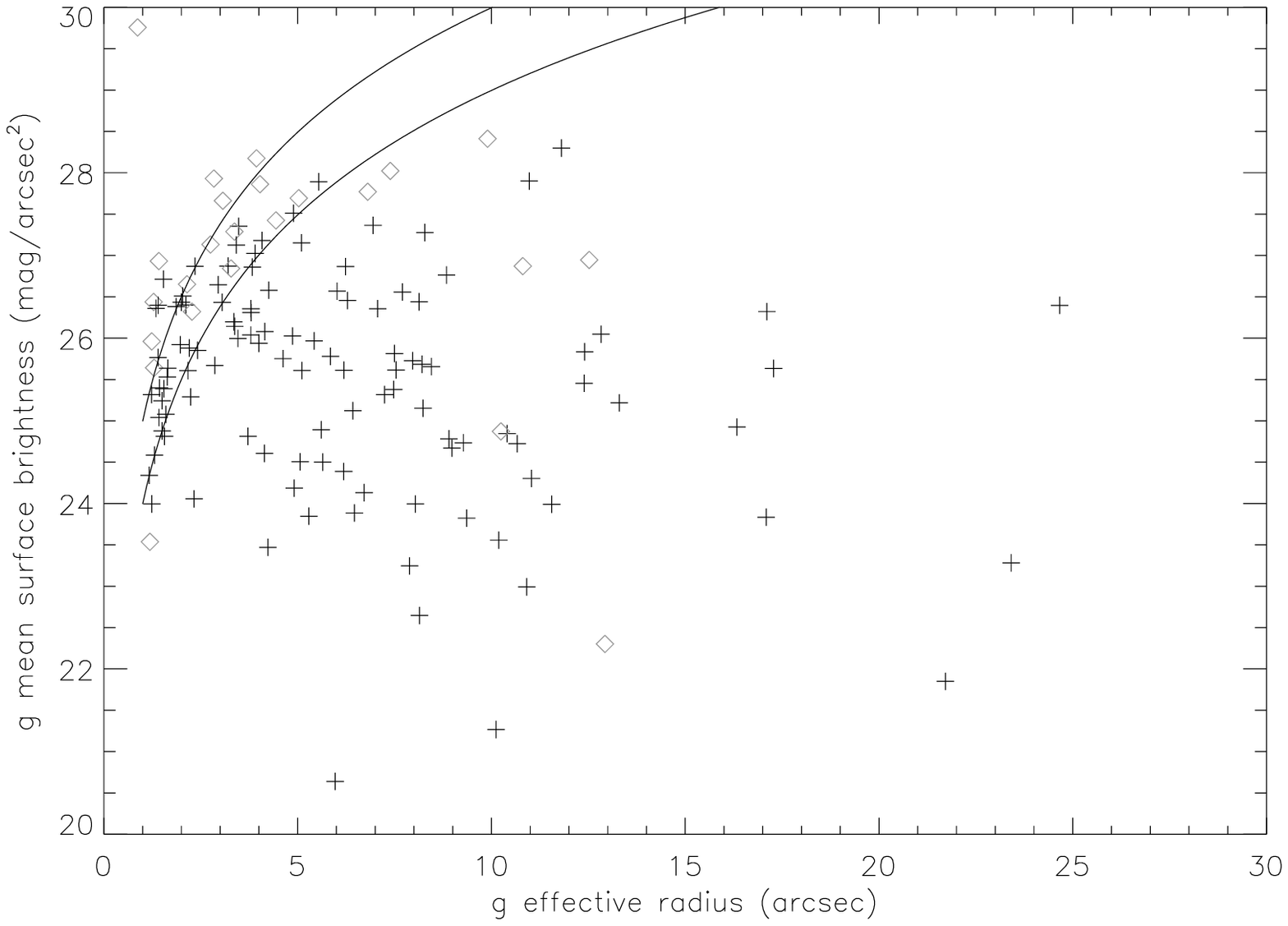}{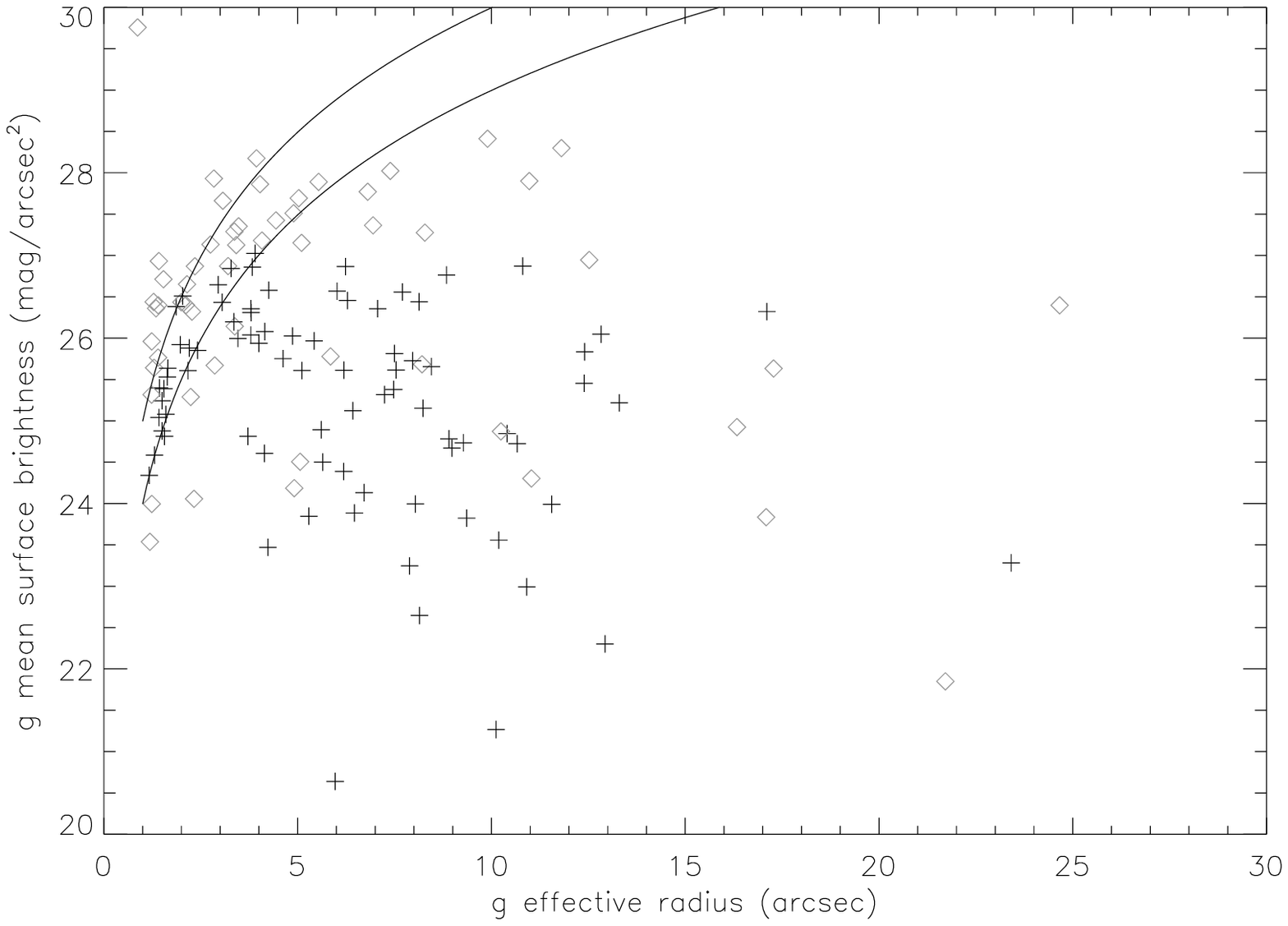}
 \put(-350,30){\bf MARSIAA}
 \put(-100,30){\bf SExtractor}
 \caption{Results for the NGVS+0+1 data: mean g-band surface brightness of the Virgo LSB galaxies as a function of  effective radius. Left panel: MARSIAA/DetectLSB. Right panel: SExtractor/DetectLSB. The solid lines correspond to g magnitudes of 22 and 23. Grey diamonds: not identified LSB galaxies. Black crosses: identified LSB galaxies. The completeness for $r_{\rm e} > 1.5''$ and $m_{\rm g} < 22$~mag is $92$\,\% for MARSIAA and $72$\,\% for SExtractor.
 \label{fig:modelgraph1}}
\end{figure}

The recovered g mean surface brightness of the identified LSB galaxies as a function of the effective radius is presented in Fig.~\ref{fig:modelgraph1}. 
The completeness for $r_{\rm e} >$ \as{1}{5}   and $m_{g} < 22$~mag is $92$\,\% for DetectLSB/MARSIAA and $72$\,\% for DetectLSB/SExtractor. A completeness of $90$\,\% is reached for sources with $r_{\rm e} > 3''$ at a mean surface brightness of $\mu_{\rm g}=27.7$~\masq and a central surface brightness of $\mu^{0}_{\rm g}=26.7$~\masq.

\begin{figure*} %16
\plottwo{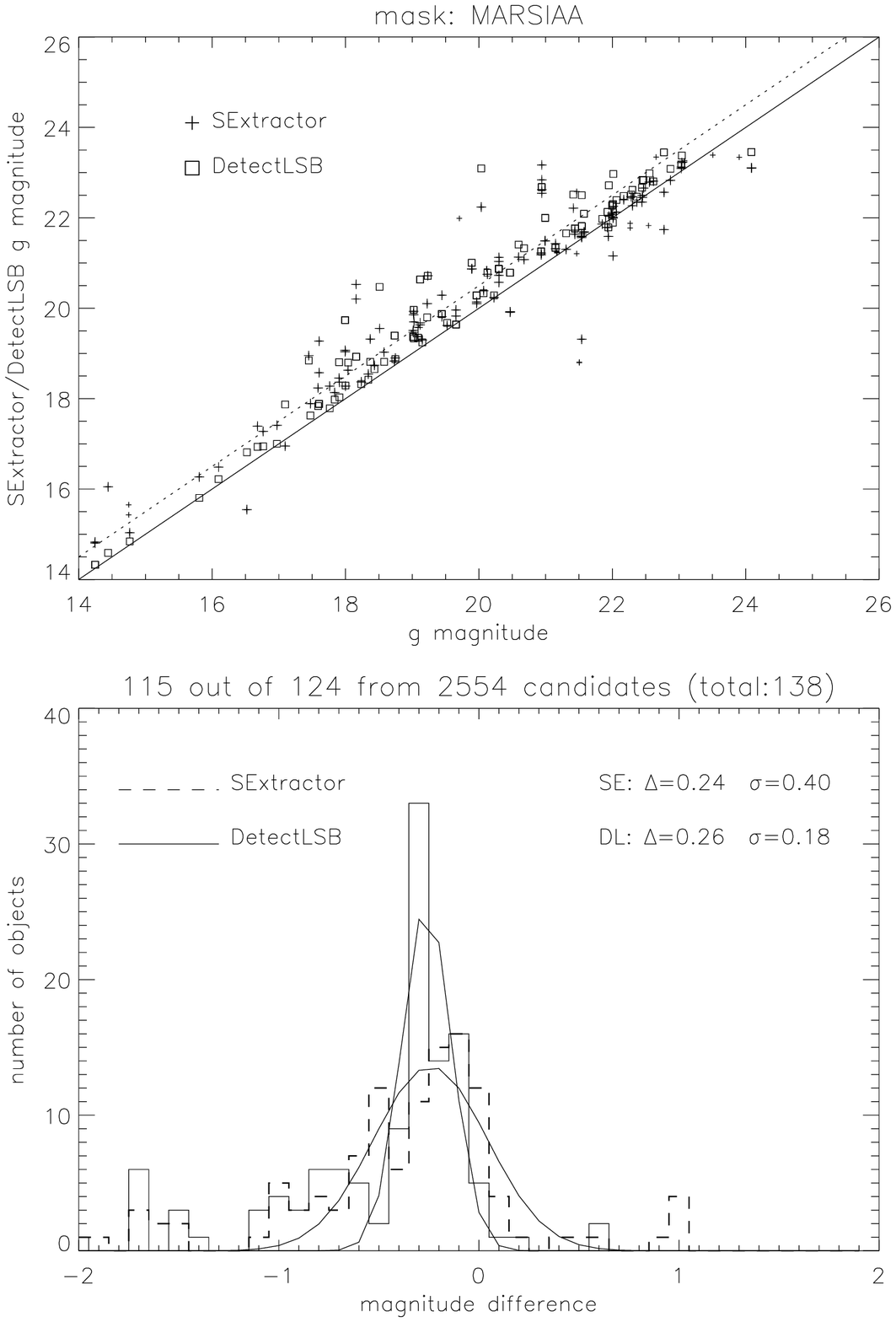}{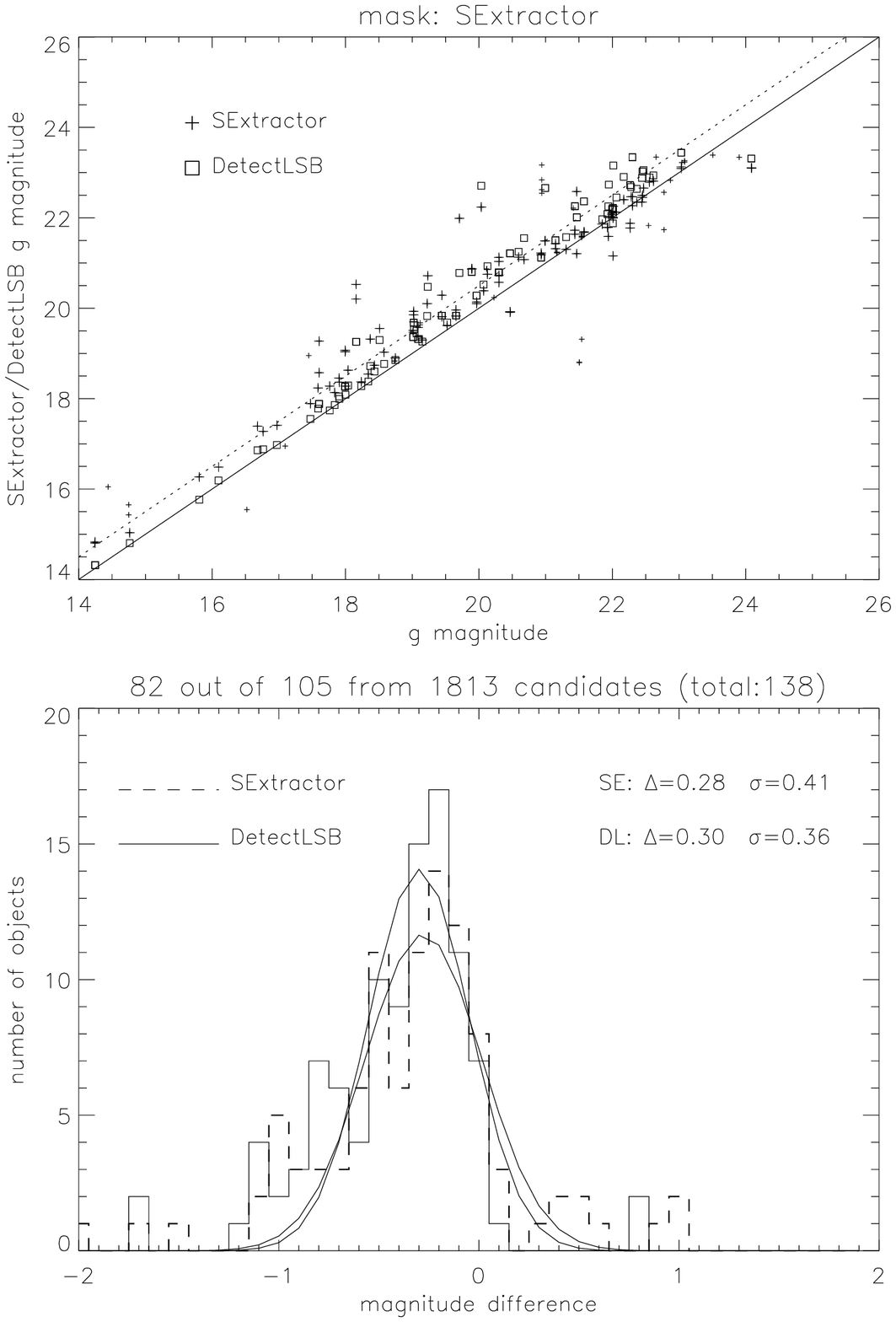}
 \caption{Results for the NGVS+0+1 data: recovered SExtractor and DetectLSB g magnitudes of the Virgo LSB galaxies extracted by the NGVS team as a function of manually fitted g magnitudes (by the NGVS team). Left panels: MARSIAA/DetectLSB. Right panels: SExtractor/DetectLSB. Upper panels: small crosses: detected but not selected LSB galaxies; open squares: DetectLSB magnitudes and half-light radii. The solid line represents equality between the recovered and the manually fitted magnitude; the dotted line has an offset of $0.5$~mag. Lower panels: distribution of the difference between the recovered and the manually fitted g band magnitudes together with Gaussian fits. 
All methods show a mean offset of $\sim 0.3$~mag from the magnitudes measured by the NGVS team, but the MARSIAA/DetectLSB method also shows a secondary peak around $-0.8$~mag.
 \label{fig:modelmag1}}
\end{figure*}

The DetectLSB and SExtractor g band magnitudes are compared to the magnitudes extracted by the NGVS team in Fig.~\ref{fig:modelmag1}.
The latter magnitudes were measured with three independent methods (2D, 1D, and non-parametric) with extremely good agreement. 
Both  methods show a magnitude offset of $0.3$~mag with respect to the values measured by the NGVS team, with a  dispersion of  $0.4$~mag.
The application of MARSIAA/DetectLSB results in a bimodal structure in  the magnitude difference distribution: a prominent peak with an offset of $0.3$~mag and a dispersion of $0.2$~mag, and a secondary peak around an offset of $-0.8$~mag. This bimodal structure is also present, although less pronounced, in the distribution of the SExtractor
magnitude differences. As expected, DetectLSB recovers less flux than a manual photometric extraction.

\begin{figure*} % 17
\plottwo{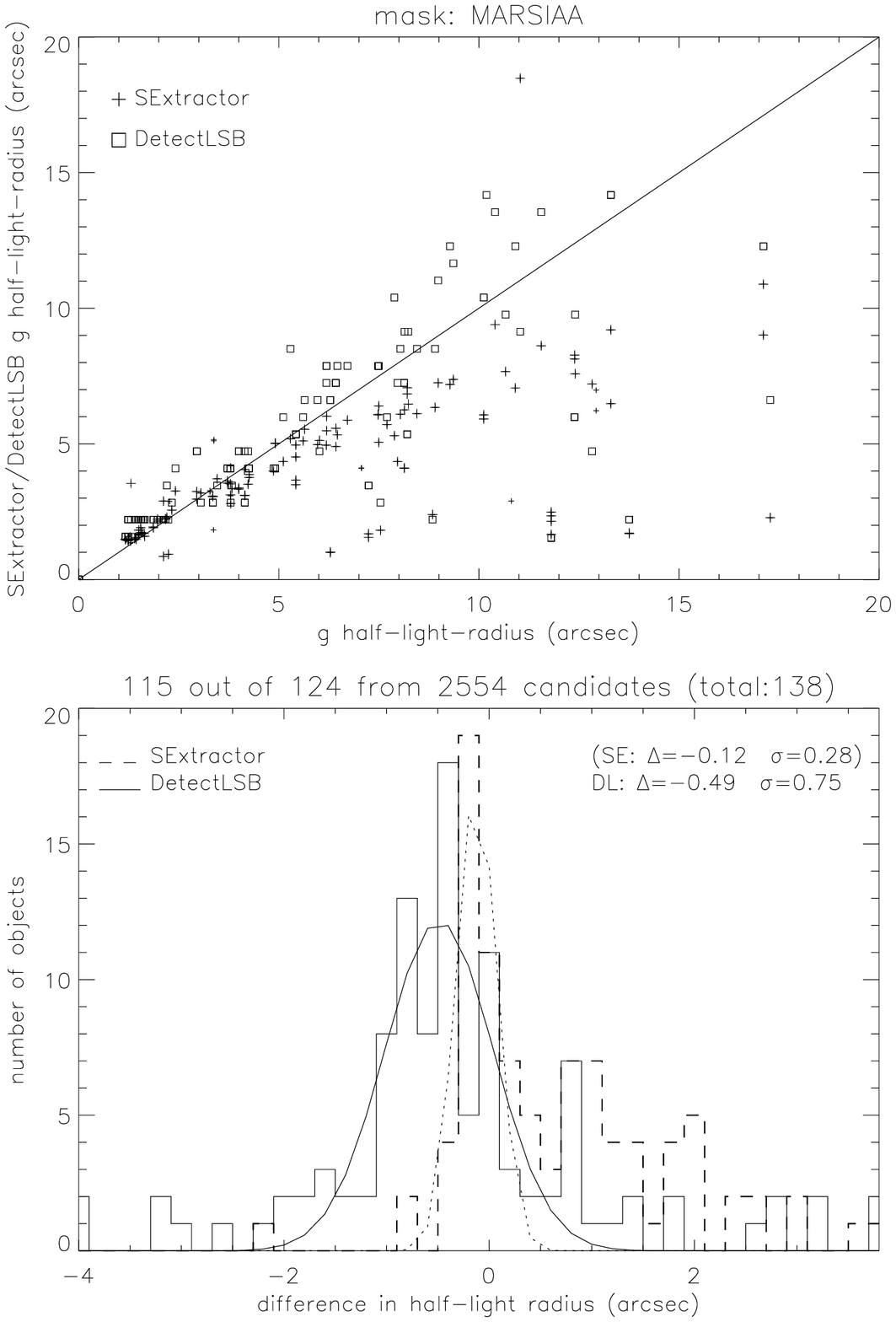}{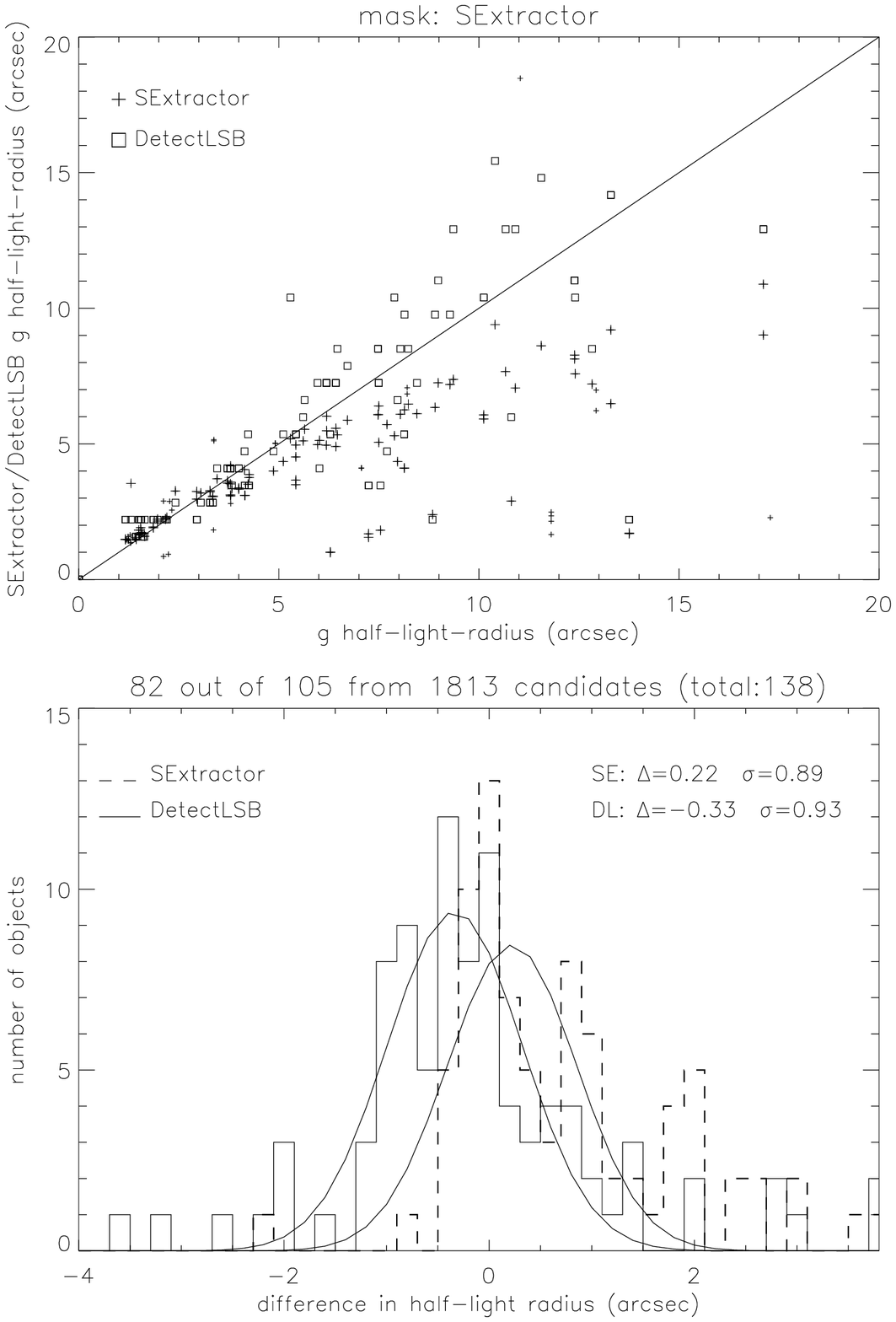}
 \caption{Results for the NGVS+0+1 data: recovered SExtractor and DetectLSB effective radii of the Virgo LSB galaxies as a function of the manually fitted effective radii. Left panels: MARSIAA/DetectLSB. Right panels: SExtractor/DetectLSB. Upper panels: small crosses: detected but not selected LSB galaxies. The solid line represents equality between the recovered and the manually fitted effective radius. Lower panels: distribution of the difference between the recovered and the manually fitted effective radius together with Gaussian fits.
The difference between the SExtractor and NGVS team radii increases with radius, whereas the DetectLSB radii have an about constant offset of $< 1''$.
 \label{fig:modelradius1}}
\end{figure*}

The DetectLSB and SExtractor half-light radii are compared to the input half-light radii in Fig.~\ref{fig:modelradius1}.
As for the mock data (Sec.~\ref{sec:ngvsmock}), the difference between the  half-light radii estimated by SExtractor and those measured by the NGVS team increases with increasing radius. On the other hand, the half-light radii estimated by DetectLSB have  an offset of less than $1''$, with a dispersion of  \as{0}{8}--$1''$.

% end of new Section

\section{Conclusions \label{sec:conclusions}}

The automated detection of low surface brightness galaxies in images has hitherto generally been accomplished by applying either (i) SExtractor, (ii) ring filtering, or (iii) matched filters. We propose a different algorithm, called MARSIAA (MARkovian Software for Image Analysis in Astronomy), which is based on multi-scale Markovian modelling. MARSIAA can be applied simultaneously to images observed in several bands (Laferte et al. 2000, Provost et al. 2004).
It segments the image into different classes according to their surface brightness and continuity, i.e., their surroundings. The number of classes is defined by the user. The result of MARSIAA is a binary map for each class. Typically, class 0 contains the noise, and class 1 and/or 2 contain the LSB structures. It is not possible to give a general S/N limit for the detection of LSB galaxies because the average surface brightness of pixels of class 0 differs from image to image, depending on the surface brightness distribution of all pixels. 
For the second step we developed an algorithm called DetectLSB to identify LSB galaxies, and remove perturbing sources recognized by MARSIAA. It simultaneously fits surface brightness profiles to the 4 quadrants of selected objects and determines the galaxy center, major axis position angle, and inclination.

MARSIAA and DetectLSB were tested on a set of simulated LSB galaxies of different signal to noise ratios and scale-lengths, surrounded by point sources of different central surface brightnesses. Applying MARSIAA to several bands simultaneously increased the detection probability. Based on our tests we are confident that we can detect LSB galaxies down to a S/N (peak surface brightness to pixel noise) ratio of $\sim 1.5$. 

To assess the  robustness of our method, MARSIAA and DetectLSB were applied to 18 INT B and I band images of the Virgo cluster (Sabatini et al. 2003), with a total of 150 megapixels covering a 1.3 square degree area. MARSIAA/DetecLSB identified about 800 potential LSB sources, of which 270 have characteristic  scale-lengths $>3''$. From this sample we classified 52 as LSB galaxies by eye. We detected all objects from the Sabatini et al. (2005) Virgo cluster LSB dwarfs catalogue that we could classify by eye as bona fide LSB galaxies, whereas applying SExtractor optimized for LSB detection fails to identify 3 of these objects. An additional 4 LSB galaxy candidates with characteristic scale-lengths $>3''$, which are not included in the Sabatini et al. (2005) catalogue, were detected by our method. 

To assess the completeness of the results of our method, MARSIAA and DetectLSB were applied to mock Virgo LSB galaxies inserted into a set of NGVS gri-band subimages of the Virgo cluster (Ferrarese et al. 2012), and to a full set of NGVS square degree gri images. Since the NGVS images are about $3$~mag deeper than the INT images, the surface density  of detected LSB structures  is much higher in the NGVS images  than  in the INT images. We developed an additional software to select LSB galaxy candidates from the DetectLSB identifications which takes advantage of the segmentation classes. It  reduces the number of Virgo LSB galaxy candidates obtained by DetectLSB to $5$-$20$\,\% of the original number, with a ratio of false positives of  about $90$\,\%. About $10$\,\% of the false positives are artifacts, the rest being background galaxies.
MARSIAA/DetectLSB identified $\sim 20$\,\% more mock Virgo LSB galaxies and $\sim 40$\,\% more Virgo LSB galaxies that were detected by eye than SExtractor/DetectLSB did. The magnitudes derived by DetectLSB are as reliable as those derived by SExtractor, with an  uncertainty  of $\sim$$0.5$~mag.
SExtractor underestimates the sizes for sources  with half-light radii in excess of $1.5''$. This offset increases with increasing half-light radius. DetectLSB overestimates the source sizes by an approximately constant offset of $\sim 0.5''$. The DetectLSB sizes are thus  approximate, but more reliable than the SExtractor  estimates. A completeness of $90$\,\% was  reached for the recovery of  sources with $r_{\rm e} > 3''$ at a mean surface brightness level of $\mu_{\rm g}=27.7$~\masq and a central surface brightness of $\mu^{0}_{\rm g}=26.7$~\masq. To weed the Virgo cluster LSB
galaxies out of the large number of LSB galaxy candidates ($\sim 90$\,\% of false positives), 
at least $1000$ objets per square degree have to be examined manually or by a smart unsupervised fitting routine, which is beyond the scope of our work.

We therefore conclude that our proposed Markovian method (MARSIAA and DetectLSB) is complementary to the application of matched filters and SExtractor, and that is has the following advantages:
\begin{enumerate}
\item
it is scale-free;
\item
it is intended to be applied simultaneously to several bands;
\item
it is well adapted for crowded regions on  the sky, where source confusion is substantial.
\end{enumerate}

\begin{acknowledgements} 
The author would like to thank the NGVS team for their support, especially L.~Ferrarese for reading the manuscript and 
giving us useful comments. 
This work has been supported by the ANR-10-BLANC-0506 program funded by the French Agence National de la Recherche.
The NGVS observations were obtained with MegaPrime/MegaCam, a joint project of CFHT and CEA/DAPNIA, at the Canada-France-Hawaii Telescope (CFHT), which is operated by the National Research Council (NRC) of Canada, the Institut National des Science de l'Univers of the Centre National de la Recherche Scientifique (CNRS) of France and the University of Hawaii.The Isaac Newton Telescope is operated on the island of La Palma by the Isaac Newton Group in the Spanish Observatorio del Roque de los Muchachos of the Instituto de Astrof{\'i}sica de Canarias.
\end{acknowledgements}

%\clearpage

%% the file, uncomment the \plotone, \plottwo, and
%% If you need a layout that cannot be achieved with \plotone or
%% \plottwo, you can invoke the graphicx package directly with the
%% \includegraphics command or use \plotfiddle. For more information,
%% please see the tutorial on "Using Electronic Art with AASTeX" in the
%% documentation section at the AASTeX Web site,
%% http://www.journals.uchicago.edu/AAS/AASTeX.
%%
%% The examples below also include sample markup for submission of
%% supplemental electronic materials. As always, be sure to check
%% the instructions to authors for the journal you are submitting to
%% for specific submissions guidelines as they vary from

%% This example uses \plotone to include an EPS file scaled to
%% 80% of its natural size with \epsscale. Its caption
%% has been written to indicate that additional figure parts will be
%% available in the electronic journal.

\appendix

\section{Multi-scale segmenation (Provost et al. 2004)}

A Hidden Markov quadtree (HMT) is an acyclic graph $G=\left(
S,L\right)$ with a set of nodes $S$ and a set of edges $L$. $S$ is
partitioned into ``scales'', \textit{i.e.;} $ S=S^{0}\cup
S^{1}\ldots \cup S^{R}$, such that $S^{R}=\{r\}$ is the root,
$S^{n}$ involves $4^{R-n}$ nodes, and $S^{0}$ is the finest scale
formed by the leaves. Each node $s$, excepts the root $r$, has a
unique predecessor, its ``parent'' $s^-$. Each node $s$, except
the ``leaves'', has four ``children'' $s^+=\{u \in S: u^-=s\}$. We
use the notation $s^{++}$ for all descendants of $s$. 
 
\begin{figure*}
 \input{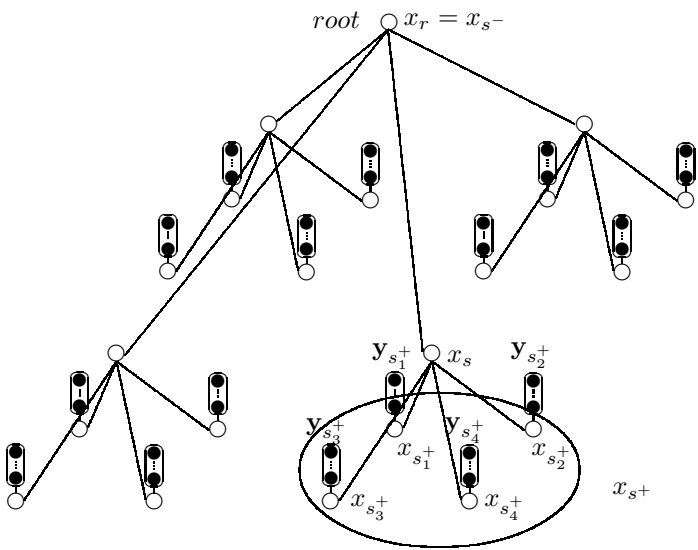}
 \caption{Example of a dependency graph corresponding to a quadtree structure on a $4 \times 4$ lattice. White circles represent labels and black circles represent multiband observations $\mathbf{y}_s$, ${s\in S}$. Each node $s$ has a unique \textit{parent} $s^-$, and four \textit{children} $s^{+}=\{s^{+}_1,\; \cdots,s^{+}_4 \}$.}
 \label{fig:quad}
\end{figure*}

The hidden process\footnote{To simplify the notation, we will denote
the discrete probability $P(X=x)$ as $P(x)$.} $X$, which assigns to each node $s \in S$ a hidden state $X_{s}$, is chosen from the label set
$\Omega =\{\omega _{1},...,\omega _{K}\}$ of the $K$ classes. $X$ is
assumed Markovian in scale, \textit{i.e.,}~:\begin{equation}
P(x^{n}| x^{k},k>n) =P(x^{n}|x^{n+1}); \;\; x^n=\{x_s:\; s\in S^n\}.\end{equation} Moreover, $X_s$, $s \in S^n$, is independent from all $X_u$, $u\in S^{n+1}$, given its parent and the inter-scale transition probability, and can be factorized in the following way:
\begin{equation}\label{factorisation}
P(x^{n}| x^{n+1}) =\prod_{s\in S^{n}}P(x_{s}| x_{s^{-}})
\end{equation}
The hidden process $X$ is called a Markov tree since it verifies
(Laferte et al. 2000):
\begin{equation}\label{HMT}
 P(x)=P(x_r) \prod_{n=0}^{R-1} \prod_{s \in S^n} P(x_s | x_{s^-})
\end{equation}
The multi-component observations $\mbox{\boldmath ${Y}$}$ are introduced at the scale $S^0$ so that each D-dimensional pixel
$\mathbf{y}_s$ is linked to the hidden state $X_{s}$ (Fig. \ref{fig:quad}). The HMT assumes $\mathbf{y}_s$ independent from
 the entire quadtree given its hidden state, which is formulated as
follows:
\begin{equation}
 P(\mathbf{y}_s | x,\mathbf{y}-\{\mathbf{y}_s\})=P(\mathbf{y}_s | x_s).
\end{equation}
Thus the probability of $\mbox{\boldmath ${Y}$}$ conditionally to
$X$ is expressed as the following product:
\begin{equation}
P(\mathbf{y}|x)=\prod_{s\in S^{0}}P(\mathbf{y} _{s}|x_{s}),
\label{eq:vraisemblance}
\end{equation}
where $\forall \; s\in S^{0}$, $P(\mathbf{y}_{s}|x_{s}=\omega
_{i})$, called a data driven term, captures the likelihood of the
observation $\mathbf{y}_{s}$ with respect to the class $\omega _{i}$. 
In the case of multidimensional Gaussian noise the covariance matrix
takes the correlation between spectral bands into account.

If no data are available at a given site $s$ in the image, \textit{i.e.,} missing or masked data, the likelihood at this site is set to $1$ (Provost et al. 2004).
Thus $P(\mathbf{y}_{s}|x_{s}=\omega _{i})$ is computed as the likelihood
of $\mathbf{y}_{s}$ of the class $\omega_{i}$.

From the assumptions above, the joint distribution
$P(\mathbf{x},\mathbf{y})$ can easily be factorized as follows :
\begin{equation}
P(\mathbf{x},\mathbf{y})=P(x_{r})\prod\limits_{s\neq
r}P(x_{s}|x_{s^{-}})\prod_{s\in S^{0}}P(\mathbf{y}_{s}|x_{s}).
\label{vraicon}
\end{equation}

The HMT parameters are:
\begin{description}
 \item[$\bullet$] $\Phi_x$, the \textit{a priori} parameters regrouping:
 \begin{description}
 \item[-] $\{\pi_i=P(x_r=\omega_i)\}_{i=1,\; \cdots,K}$ the probabilies at the root,
 \item[-] $\{a_{ij}=P(x_s=\omega_j | x_{s^-}=\omega_i)\}_{i,j=1,\; \cdots,K}$ the parent/child transition probabilities,
 \end{description}
 \item[$\bullet$] $\Phi_y$, the parameters of the likelihoods $\{P(.|x_{s}=\omega _{i})\}_{i=1,\;
 \cdots,K}$.
\end{description}

In the pdf version, there is text here on "Algorithmus 1"; I do not see that in the Latex file. Algorithmus = Algorithm

One of the interests of this model is the possibility of computing exactly the \textit{posterior} marginals $P(x_{s} | \mathbf{y})$ and $%
P(x_{s},x_{s}^{-} | \mathbf{y})$ at each node~$s$ in two passes on the quadtree (Algorithm \ref{table:up_down}).

\begin{algorithm*}[htd]
 \begin{algorithmic}
 \STATE{$\bullet$ Evaluation of the partial
 posterior marginals at the bottom of the quadtree : $$\forall s\in
S^0,\; P(x_{s}=\omega_i/\mathbf{y}_{s^{++}})=
P(x_{s}=\omega_i/\mathbf{y}_{s})=\frac{ P(x_{s}=\omega_i)
P(\mathbf{y}_{s}/x_{s}=\omega_i)}{\sum_{\omega_j} P(x_{s}=\omega_j)
P(\mathbf{y}_{s}/x_{s}=\omega_j)},$$

where $P(x_{s}=\omega_i)$ is recursively evaluated through a
top-down pass, given the \textit{ prior} probability
$P(x_{r}=\omega_i)=\pi_i$ as follows :

\FOR { $n= R-1, \cdots, 0$} \FORALL {$s \in S^n$} \STATE{
$P(x_{s}=\omega_i)=\sum_{\omega_j}
P(x_{s}=\omega_i/x_{s^-}=\omega_j)P(x_{s^-}=\omega_j)$} \ENDFOR
\ENDFOR
 }.

\STATE{$\bullet$ \textbf{Upward pass } :\\
\FOR { $n= 1, \cdots, R$} \FORALL {$s \in S^n$}
\STATE{$P(x_{s}=\omega_i/\mathbf{y}_{s^{++}})=\frac{1}{Z}
P(x_{s}=\omega_i) \prod_{t\in s^+} \sum_{\omega_j}
\frac{a_{ij}P(x_{t}=\omega_j/\mathbf{y}_{t^{++}})}{P(x_{t}=\omega_j)}$}
\ENDFOR \ENDFOR

where $a_{ij}=P(x_{t}=\omega_j/x_{t^-}=\omega_i)$ is the parent/child transition probability\\ and $Z$ is a normalizing factor such that $\sum_{\omega_i}
P(x_{s}=\omega_i/\mathbf{y}_{s^{++}})=1$. Note that at the top of
quadtree we obtain $P(x_{r}=\omega_i/\mathbf{y})$}

\STATE{$\bullet$ \textbf{Downward pass } :\\
\FOR { $n= R-1, \cdots, 0$} \FORALL {$s \in S^n$}
\STATE{$$P(x_{s}=\omega_j,x_{s^-}=\omega_i/\mathbf{y})=
P(x_{s^-}=\omega_i/\mathbf{y})\frac{P(x_{s}=\omega_j/\mathbf{y}_{s^{++}})a_{ij}P(x_{s^-}=\omega_i)/P(x_{s}=\omega_j)}{\sum_{\omega_l}
P(x_{s}=\omega_l/\mathbf{y}_{s^{++}})a_{il}P(x_{s^-}=\omega_i)/P(x_{s}=\omega_l)},$$

$$P(x_{s}=\omega_j/\mathbf{y})=
\sum_{\omega_i}P(x_{s}=\omega_j,x_{s^-}=\omega_i/\mathbf{y})$$ }
\ENDFOR \ENDFOR }
 \end{algorithmic}
 \caption{Two passes on the quadtree for \textit{posterior}
computation given HMT parameters $\{\Phi_x,\Phi_y\}$.}
\label{table:up_down}
 \end{algorithm*}

The EM algorithm used for the estimation of the \textit{a priori}
parameters $\Phi_x$, leads to an iterative procedure with the
followings updates (Flitti et al. 2005, Provost et al. 2004) :

\begin{eqnarray}\label{appparam}
 \nonumber a_{ij}^{[c+1]} &=& \frac{\sum_{s \in S^n,\;n \neq r} p^{[c]}(x_s=\omega_j,x_{s^-}=\omega_i | \mathbf{y})}{\sum_{s \in S^n,\;n \neq r} p^{[c]}(s_{s^-}=\omega_i | \mathbf{y})} \\
 \pi_i^{[c+1]} &=& p^{[c]}(x_r=\omega_i | \mathbf{y})
\end{eqnarray}
where $[c]$ stands for the current iteration and
$p^{[c]}(x_s=\omega_i | \mathbf{y})$ and
$p^{[c]}(x_s=\omega_j,x_{s^-}=\omega_i | \mathbf{y})$ are computed
by way of the two passes of Algorithm \ref{table:up_down} using
the current parameters. 

When converged, i.e., when the difference between successive
updates is small enough or the maximum number of iteration is
reached, the Marginal \textit{a Posteriori} Mode criterion (MPM)
is used to obtain the segmentation map:
\begin{equation}\label{MPM}
\forall \; s \in S^0, \;\hat{x}_s = \arg \max_{x_s \in \Omega}
p(x_s | \mathbf{y})
\end{equation}

\clearpage

\begin{table}
 \caption{Rejection statistics for three independent visual inspections (by SS., WvD., BV.) of 10 INT images.}
 \label{tab:rejection}
 \[
 \begin{tabular}{lllll}
 \hline
	 number of sources & 479 & 497 & 414 \\
	 rejected sources & 250 & 293 & 279 \\
	 multiple & 113 & 153 & 193 \\
	 noise & 105 & 70 & 55 \\
	 stellar halo & 25 & 46 & 6 \\
	 stripe & 0 & 0 & 6 \\
	 background & 0 & 3 & 0\\
	 track & 0 & 3 & 3 \\
	 saturated & 3 & 0 & 0 \\
	 \hline &
 \end{tabular}
 \]
\end{table}

\begin{table*}
 \caption{Comparison of LSB galaxy detections with different methods.}
 \label{tab:comparison}
 \[
 \begin{tabular}{llllll}
 \hline
	 [SDV2005]$^{\rm a}$ & other name$^{\rm b}$ & M+D$^{\rm c}$ & SE$^{\rm d}$ B & SE$^{\rm e}$ I & comment\\
	 \hline
	 0 & & y & n & n & \\
	 1 & & y & n & n \\
	 2 & & n & n & n & very faint\\
	 3 & t322 & n & y & n & confused/double\\
	 31 & t243 & y & y & y & \\
	 32 & t171 & y & y & y & \\
	 33 & & y & y & y & \\
	 34 & t388 & y & y & y & \\
	 35 & t411 & y & n & n & \\
	 36 & t409 & y & y & y & \\
	 37 & VCC1464 & y & y & y & \\
	 38 & & n & n & n & double\\
	 39 & & n & n & n & very faint\\
	 40 & t224 & y & y & y & \\
	 41 & & n & n & y & fuzzy object in I, confused\\
	 42 & t373 & y & y & y & \\
	 43 & VCC1538 & y & y & y & \\
	 44 & VCC1680 & y & y & y & \\
	 45 & & n & n & n & very faint\\
	 47 & VCC1681 & y & y & y & \\
	 48 & VCC1729 & y & y & y & \\
	 49 & VCC1754 & y & y & y & \\
	 50 & t231 & y & y & y & \\
	 51 & t221 & y & y & y & \\
	 52 & & n & n & n & faint double/triple\\
	 53 & t404 & y & y & y & \\
	 54 & t142 & y & y & y & \\
	 57 & t322 & n & n & n & border\\
%	 & t310 & n & n & n & nothing in B \\
%	 & t421 & n & y & y & LSB, I, cluster 11, R$\sim2.8''$\\
	 \hline &
 \end{tabular}
 \]
 \begin{list}{}{}
 \item[$^{\rm{a)}}$] Sabatini et al. (2005) matched filters detection
 \item[$^{\rm{b)}}$] VCC: Bingelli et al. (1985), t: Trentham \& Hodgkin (2002)
 \item[$^{\rm{c)}}$] MARSIAA+DetectLSB, simultaneously on B and I bands
 \item[$^{\rm{d)}}$] SExtractor (Bertin \& Arnouts 1996) on B band image only
 \item[$^{\rm{d)}}$] SExtractor on I band image
 \end{list}
\end{table*}

\begin{table*}
 \caption{MARSIAA LSB galaxy candidate detections.}
 \label{tab:lsbtable}
 \[
 \begin{tabular}{lllllllllllll}
 \hline
	 [VPC & [SDV & RA & DEC & $\mu_{0}^{\rm B}$ & $r_{\rm d}^{\rm B}$ & $\mu_{0}^{\rm I}$ & $r_{\rm d}^{\rm I}$ & $\nabla_{\rm B-I}$ & $m_{\rm B}$ & $m_{\rm I}$ & B-I & back \\
	 2011] & 2005] & (deg) & (deg) & (mag & ($''$) & (mag & ($''$) & & (mag) & (mag) & (mag) & ground \\
	 & & & & arsec$^{-2}$) & & arsec$^{-2}$) & & & & & & \\
	 \hline
	 1* & & 187.79100 & +10.9489 & 26.0 & - & 24.0 & 4 & -1.3 & 23.0 & 18.6 & 4.3 & \\
	 2* & & 188.07269 & +10.9968 & 25.8 & 4 & 25.0 & 2 & -1.3 & 21.9 & 21.6 & 0.3 & \\
	 3 & 33 & 187.86490 & +10.9357 & 23.2 & 3 & 21.9 & 3 & 0.40 & 18.4 & 17.1 & 1.3 & \\
	 4 & 32 & 187.90379 & +11.0077 & 23.7 & 8 & 22.6 & 9 & 0.50 & 17.7 & 16.4 & 1.2 & \\
	 5 & 0 & 188.06281 & +10.9383 & 25.7 & 7 & 24.2 & 6 & -0.5 & 19.9 & 18.1 & 1.7 & \\
	 6 & 31 & 188.00140 & +11.0232 & 24.1 & 5 & 22.7 & 5 & 0.30 & 18.4 & 17.0 & 1.4 & \\
	 7 & 34 & 187.73950 & +11.0879 & 24.6 & 4 & 23.5 & 3 & -0.8 & 19.5 & 18.8 & 0.7 & \\
	 8 & & 187.72220 & +10.9974 & 24.9 & 3 & 22.2 & 2 & -1.3 & 20.8 & 18.3 & 2.4 & bg \\
	 9 & 35 & 187.72411 & +10.9121 & 25.6 & 10 & 24.5 & 5 & -0.8 & 19.3 & 19.1 & 0.2 & \\
	 10 & & 187.69209 & +11.0130 & 22.5 & 5 & 21.1 & 4 & -0.7 & 17.1 & 16.1 & 1.0 & bg \\
	 11* & & 188.59621 & +10.9634 & 24.3 & 9 & 23.2 & 9 & -0.4 & 16.5 & 15.3 & 1.2 & \\
	 12 & 36 & 188.43871 & +10.8729 & 25.3 & 7 & 24.1 & 7 & -0.4 & 19.4 & 18.2 & 1.1 & \\
	 13 & & 188.42810 & +11.0229 & 22.8 & 4 & 21.7 & 3 & -0.6 & 17.5 & 16.7 & 0.7 & bg \\
	 14 & 37 & 188.22459 & +11.1914 & 23.7 & 9 & 22.7 & 10 & 0.30 & 17.1 & 16.2 & 0.9& \\
	 15 & 40 & 188.18730 & +10.9498 & 24.1 & 7 & 22.9 & 6 & -0.6 & 17.8 & 16.8 & 0.9 & \\
	 16 & & 188.23900 & +11.0278 & 22.6 & 6 & 20.5 & 4 & -0.9 & 16.5 & 15.2 & 1.3 & bg \\
	 17 & & 188.15320 & +11.0164 & 22.9 & 3 & 20.3 & 2 & -0.9 & 17.8 & 15.8 & 1.9 & bg \\
	 18 & 1 & 188.38789 & +11.0877 & 23.3 & 3 & 20.8 & 3 & -0.8 & 18.1 & 15.6 & 2.5&bg \\
	 19 & 42 & 188.67410 & +11.1428 & 24.6 & 5 & 23.2 & 5 & 0.50 & 19.0 & 17.3 & 1.7 & \\
	 20 & 43 & 188.52530 & +11.0542 & 25.2 & 5 & 23.7 & 5 & 0.40 & 19.7 & 18.6 & 1.1 & \\
	 21 & & 188.62250 & +11.1551 & 23.6 & 3 & 22.3 & 2 & -0.8 & 18.7 & 17.9 & 0.8 & bg \\
	 22* & & 189.09241 & +10.8617 & 26.0 & 3 & 24.1 & 3 & -0.5 & 21.7 & 19.4 & 2.3 & \\
	 23 & 44 & 189.15320 & +10.9912 & 24.2 & 4 & 22.7 & 4 & -0.4 & 19.0 & 17.5 & 1.4 & \\
	 24 & & 188.95110 & +10.8989 & 22.7 & 4 & 20.7 & 2 & -1.0 & 17.6 & 16.3 & 1.3 & bg \\
	 25 & & 189.13519 & +10.9072 & 22.7 & 3 & 20.4 & 3 & 0.40 & 17.9 & 15.7 & 2.2 & bg \\
	 26* & & 189.03870 & +10.8728 & 24.2 & 3 & 23.3 & 3 & 0.40 & 20.0 & 18.4 & 1.6 & \\
	 27 & 42 & 188.67400 & +11.1428 & 24.4 & 4 & 23.0 & 4 & -0.5 & 19.1 & 17.8 & 1.3 & \\
	 28 & 47 & 189.15630 & +11.1535 & 24.0 & 6 & 22.2 & 5 & -0.5 & 17.9 & 16.3 & 1.6 & \\
	 29* & & 188.90450 & +11.0620 & 25.1 & 3 & 23.6 & 3 & 0.40 & 21.0 & 19.4 & 1.5 & \\
	 30* & & 188.83600 & +11.1119 & 25.4 & 4 & 23.7 & 3 & -0.6 & 20.4 & 19.1 & 1.3 & \\
	 31* & & 188.90450 & +11.0618 & 25.1 & 3 & 23.2 & 2 & -1.1 & 20.9 & 19.2 & 1.7 & \\
	 32 & 47 & 189.15620 & +11.1536 & 23.9 & 6 & 22.2 & 6 & -0.4 & 17.8 & 16.2 & 1.6 & \\
	 33 & & 189.42470 & +10.9981 & 23.0 & 3 & 20.9 & 2 & -0.6 & 18.4 & 16.5 & 1.9 & bg \\
	 34 & 48 & 189.44209 & +10.9852 & 24.1 & 9 & 22.4 & 8 & -0.4 & 17.9 & 16.2 & 1.7&bg\\
	 35 & & 189.29970 & +11.2058 & 24.8 & 3 & 21.7 & 2 & -1.2 & 20.5 & 17.9 & 2.6 & bg \\
	 36 & & 189.25011 & +10.9344 & 24.0 & 3 & 21.1 & 1 & -1.4 & 19.4 & 17.3 & 2.1 & bg \\
	 37 & 49 & 189.57140 & +11.1807 & 25.1 & 14 & 23.7 & 13 & -0.4 & 17.9 & 16.5 & 1.4&\\
	 38 & & 189.45200 & +11.1823 & 26.4 & 9 & 23.5 & 2 & -2.1 & 22.5 & 19.9 & 2.6 & \\
	 39 & 50 & 189.47980 & +11.1485 & 25.2 & 11 & 23.7 & 9 & -0.6 & 18.3 & 16.9 & 1.3 & \\
	 40 & & 189.48000 & +11.1184 & 23.0 & 3 & 21.7 & 3 & -0.7 & 18.1 & 17.0 & 1.0 & bg \\
	 41 & & 189.48019 & +11.1482 & 25.1 & 9 & 23.7 & 8 & -0.5 & 18.4 & 17.1 & 1.2 & \\
	 42 & 51 & 190.07660 & +10.9965 & 24.2 & 7 & 22.7 & 6 & -0.5 & 18.1 & 16.7 & 1.3 & \\
	 43* & & 189.90739 & +10.9758 & 22.5 & 6 & 20.8 & 5 & -0.4 & 16.6 & 15.0 & 1.5 & \\
	 44 & 51 & 190.07651 & +10.9965 & 24.2 & 7 & 22.7 & 6 & -0.5 & 18.1 & 16.7 & 1.3 & \\
	 45* & & 189.95599 & +10.9650 & 23.8 & 3 & 22.3 & 2 & -0.8 & 19.4 & 18.1 & 1.2 & \\
	 46 & & 190.00470 & +10.9721 & 23.4 & 3 & 22.1 & 3 & -0.7 & 18.3 & 16.8 & 1.4 & bg \\
	 47 & & 189.74170 & +10.9396 & 22.8 & 4 & 20.9 & 3 & -0.8 & 17.6 & 16.1 & 1.5 & bg \\
	 48 & & 189.74820 & +10.9495 & 22.7 & 3 & 21.0 & 2 & -0.7 & 18.0 & 16.6 & 1.4 & bg \\
	 49 & 53 & 190.15480 & +11.1240 & 24.5 & 3 & 23.0 & 3 & -0.5 & 19.9 & 18.4 & 1.4 & \\
	 50 & & 190.30020 & +10.9341 & 22.6 & 6 & 21.5 & 6 & -0.3 & 16.2 & 15.3 & 0.9 & bg \\
	 51 & 54 & 190.36411 & +11.1460 & 23.2 & 6 & 21.7 & 6 & -0.3 & 17.0 & 15.4 & 1.6 & \\
	 52 & 54 & 190.36411 & +11.1461 & 23.3 & 6 & 21.7 & 6 & 0.30 & 17.1 & 15.5 & 1.6 & \\
	 \hline
 \end{tabular}
 \]
\end{table*}

\end{document}